\documentclass[a4paper,10pt]{article}

\usepackage{comment}
\usepackage{float}
\usepackage{amsmath}
\usepackage[pdftex]{color,graphicx,hyperref}
\usepackage{amsmath,amsfonts,amssymb}
\usepackage[margin=1in]{geometry}
\usepackage{setspace}
\usepackage{booktabs}
\usepackage[labelfont=bf,labelsep=period]{caption}
\usepackage{subcaption}
\usepackage[utf8]{inputenc}
\usepackage[affil-it]{authblk}
\usepackage{amsthm}
\usepackage{pdfpages}
\usepackage[square,numbers]{natbib}

\title{Distinguishing subsampled power laws from other heavy-tailed distributions}

\author[1*]{Silja Sormunen}
\author[2]{Lasse Leskel\"{a}}
\author[1]{Jari Saram\"{a}ki}

\affil[1]{\small{Department of Computer Science, Aalto University, 00076 Espoo, Finland}}
\affil[2]{\small{Department of Mathematics and Systems Analysis\\ Aalto University\\ 00076 Espoo, Finland}}
\affil[*]{\small{Corresponding author's email: silja.sormunen@aalto.fi}}
\date{}

\begin{document}

\maketitle




\begin{abstract}
	Distinguishing power-law distributions from other heavy-tailed distributions is challenging, and this task is often further complicated by subsampling effects. In this work, we evaluate the performance of two commonly used methods for detecting power-law distributions -- the maximum likelihood method of Clauset \emph{et al.}~and the extreme value method of Voitalov \emph{et al.}~-- in distinguishing subsampled power laws from two other heavy-tailed distributions, the lognormal and the stretched exponential distributions. We focus on a random subsampling method commonly applied in network science and biological sciences.  In this subsampling scheme, we are ultimately interested in the frequency distribution of elements with a certain number of constituent parts -- for example, species with $k$ individuals or nodes with $k$ connections -- and each part is selected to the subsample with an equal probability. We investigate how well the results obtained from low-subsampling-depth subsamples generalize to the original distribution.  Our results show that the power-law exponent of the original distribution can be estimated fairly accurately from subsamples, but classifying the distribution correctly is more challenging. The maximum likelihood method falsely rejects the power-law hypothesis for a large fraction of subsamples from power-law distributions. While the extreme value method correctly recognizes subsampled power-law distributions with all tested subsampling depths, its capacity to distinguish power laws from the heavy-tailed alternatives is limited. However, these false positives tend to result not from the subsampling itself but from the estimators' inability to classify the original sample correctly. In fact, we show that the extreme value method can sometimes be expected to perform better on subsamples than on the original samples from the lognormal and the stretched exponential distributions, while the contrary is true for the main tests included in the maximum likelihood method.
\end{abstract}


\section{\label{sec:level1}Introduction}
 
Power-law distributions have frequently been observed in both natural and artificial systems. They naturally emerge by mechanisms such as preferential attachment \cite{Barabasi_Albert_1999} and restarts in telecom and queueing networks \cite{Jelenkovic_Tan_2007,Asmussen_Fiorini_Lipsky_Rolski_Sheahan_2008}, and they also serve to explain other complex phenomena such as fractional Gaussian noises \cite{Kaj_Leskela_Norros_Schmidt_2007} and small-world phenomena \cite{VanDerHofstad_Hooghiemstra_Znamenski_2007}. The apparent ubiquity of power laws has been taken to indicate a universal self-organizing mechanism at play, and power laws have acquired a reputation as a hallmark of complex systems. However, there is no universally accepted method for identifying power laws, and their ubiquity has been argued to result from lacking statistical testing rather than their actual universality \cite{broido}. The debate has been especially heated in the field of network science, where networks with a power-law degree distribution, the so-called scale-free networks, play a prominent role. For these networks, the probability that a randomly chosen node has $k$ connections to other nodes varies as a power of the degree $k$. Using the maximum likelihood methods presented in Ref.~\cite{clauset}, Broido and Clauset concluded that -- contrary to the widespread belief -- scale-free networks are rare \cite{broido}. A power law was required to hold only for the largest degrees: 
\begin{eqnarray}
    &P(k) = \frac{1}{\zeta(\alpha,k^{\prime}_{\mathrm{min}})}k^{-\alpha}, \ \text{for degrees}\ k \geq k^{\prime}_{\mathrm{min}} \geq 1,
\end{eqnarray}

\noindent
where $\alpha > 1$ is the tail exponent, $k^{\prime}_{\mathrm{min}}$  is the smallest degree for which the power law holds, and $\zeta(\alpha,k^{\prime}_{\mathrm{min}})$ is the Hurwitz zeta function which normalizes the degree distribution so that $\sum_{k=k^{\prime}_{\mathrm{min}}}^{\infty} P(k) = 1$. In contrast to this, however, Voitalov \emph{et al.}~\cite{voitalov} found more evidence of scale-free networks using a method based on extreme value theory. In their work, the definition of power-law distribution was extended to include all regularly varying distributions, defined as distributions whose complementary cumulative distribution function approaches a power law asymptotically in the tail while deviating arbitrarily from a pure power law for smaller degrees. In line with this, Serafino \emph{et al.}~\cite{serafino} found many empirical degree distributions to satisfy the scale-free hypothesis when finite-size effects were taken into account with tools from statistical physics. Note that while these articles use the terminology of network science, the methods can equally well be applied to frequency distributions in any other domain of science.

In summary, distinguishing power-law distributions from other heavy-tailed distributions has proven to be challenging, and this task is further complicated if the data is \emph{subsampled}. Often only a part of the system can be observed, and depending on the subsampling strategy, the subsample represents the original system more or less accurately. For example, if we sample nodes of a network with equal probability and record their degrees, the subsampling is trivial in the sense that the degree distribution stays unchanged. However, this kind of unbiased subsampling is often not possible; we might, for example, sample each node with equal probability but only be able to observe the connections between the chosen nodes, in which case the observed degree distribution does not faithfully reflect that of the whole system. In such cases, making inferences from the subsample without further consideration might lead to erroneous conclusions.

In this work, we assess how reliably two state-of-the-art methods for recognizing power-law distributions  -- that of Clauset \emph{et al.}~\cite{clauset,broido} and Voitalov \emph{et al.}~\cite{voitalov} -- determine whether subsampled data originates from a power-law distribution. We focus on a random subsampling method where the probability of observing a node depends on its degree. In network science, this method is known as  the incident subgraph sampling strategy, where each edge is included in the subsample with probability $\pi$ together with the nodes that it connects. The degree distribution of the subsamples is given by
\begin{eqnarray}
    &P_s(k) = \sum_{i \geq k}^{\infty} P(i) \binom{i}{k}\pi^k(1-\pi)^{i-k},
\label{eq:subsampling}
\end{eqnarray}

\noindent
where $P(k)$ denotes the degree distribution of the original network. This subsampling strategy is equivalent to selecting each node with a probability linear in degree. The same strategy can be applied to frequency distributions arising in other contexts. In biodiversity studies, for example, $P(k)$ might represent the fraction of species with $k$ individuals, and each individual would subsequently be picked to the subsample with probability $\pi$. In general, this subsampling method is common in situations where observing the whole system or population is impossible 
due to the sheer size of the system, such as when recording the size distribution of neuronal avalanches in the brain \cite{levina}, assessing relative species abundance in an ecological community \cite{shimadzu} or investigating the diversity of cells in immunological studies \cite{heikkila}. In such cases, it is crucial to know to what extent the results obtained for a subsample can be generalized to the original distribution.

For power laws, generalizing the results to the original distribution is not straightforward. 
Stumpf \emph{et al.}~\cite{stumpf_randomgraphs} have argued that the degree distribution of a scale-free network is not closed under the subsampling strategy described by Eq.~2, meaning that the degree distribution of the subsampled network and that of the whole network do not belong to the same family of probability distributions. The same applies to other heavy-tailed distributions such as the lognormal and the stretched exponential distributions \cite{stumpf_randomgraphs}, which are notoriously difficult to distinguish from power laws and have commonly been used as alternatives to the power-law hypothesis in the previous literature (see for example Refs.~\cite{malevergne2003, malevergne, montebruno, foss}). For power-law distributions, deviation from the original power law grows larger as the subsampling depth decreases or the power-law exponent increases \cite{stumpf_randomgraphs}. However, the form of the distribution is mostly affected for small degrees, and the tail of the subsampled distribution still approaches the original power law asymptotically for $k \gg 1$ \cite{stumpf_scalefree}.  This finding is fruitfully exploited by Levina \emph{et al.}~\cite{levina}, who propose a method for differentiating between power-law and exponential distributions by further subsampling the data -- a subsample itself -- and scaling the subsampled sequences in a way that collapses the scaled tails of the subsamples to the original power law. 

In general, methods considering only the tail of a distribution should -- in theory -- continue to work on subsampled power-law distributions, which still belong to the larger class of regularly varying distributions. However, we do not currently know how incident subgraph sampling affects the separability of other heavy-tailed distributions from power-law distributions. Subsampling might distort other distributions to resemble power laws; for example Han \emph{et al.}~\cite{han} have previously observed that when the subsampling scheme consists of selecting a fraction $p$ of the nodes and a fraction $q$ of those nodes neighbors, subsamples from networks with exponential, truncated normal and Poisson degree distributions can mimic power-law-like behavior, when the resemblance to power law is assessed based on the degree of linearity between logarithms of the degree $k$ and the number of nodes with degree $k$.

 In this work, we evaluate how reliably the methods of Clauset \emph{et al.}~\cite{clauset,broido} and Voitalov \emph{et al.}~\cite{voitalov} succeed in distinguishing subsampled power-law distributions from two other types of heavy-tailed distributions -- namely the lognormal and the stretched exponential distributions -- when the above-described incident subgraph sampling strategy is applied to subsamples from simulated degree distributions. We use the term stretched exponential distribution to refer to the subclass of Weibull distributions with $\beta \in (0,1)$ to maintain consistency with nomenclature in our core references. For convenience, we use the name maximum likelihood (ML) method to refer to the power-law hypothesis test by Broido and Clauset \cite{broido} based on the methods of Clauset \emph{et al.}\ \cite{clauset}, and the name extreme value (EV) method to refer to the method of Voitalov \emph{et al}. As heavy-tailed distributions are by definition heavier than any exponential distribution, we assess the methods' performance on discrete exponential distributions (i.e.\ geometric distributions) as well. Our analysis is restricted to these two methods because their implementation is readily available, excluding, e.g., the finite size scaling method of Ref.~\cite{serafino}. It is also worth noting that this finite size scaling method uses the tools of the ML method to determine how large a fraction of the distribution's tail is to be considered in the analysis, and therefore its performance depends partly on how well this estimation succeeds.

Our results show that the power-law exponent can be estimated fairly accurately from simulated subsamples of power-law distributions, but the classification of the distribution's type should be taken with caution. Finally, we show that subsampling affects the performance of the two methods differently: while the EV method can in some cases be expected to perform better on subsampled data than on the original distribution, the opposite applies to the main tests included in the ML method.

\section{\label{sec:estimators}Estimators}

We start by briefly presenting the maximum likelihood (ML) and the extreme value (EV) methods; the reader already familiar with these can move straight to Section \ref{sec:simulating_subsampling}. Note that we have chosen to use these names for the sake of convenience, and they do not necessarily capture the essence of the methods; the EV method, for example, 
incorporates estimators that are based on the maximum likelihood approach, while the ML method contains additional tests not based on maximum likelihood estimation.

\subsection{\label{sec:estimators_clauset}Maximum likelihood (ML) method}

The ML method of Refs.~\cite{broido,clauset} for assessing whether a sample originates from a power-law distribution starts with estimating the optimal values of the start of the power law and the corresponding power-law exponent. The rationale behind not necessarily including the entire sample in the analysis is that many empirical distributions are expected to follow a power law only for large values of $k$  \cite{clauset}. In the following, we denote the true start of the power law (a property of the distribution) with $k^\prime_\mathrm{min}$. We use $\hat{k}_\mathrm{min}$ to refer to the best estimate of $k^\prime_\mathrm{min}$ produced by the ML method; only data points $k \geq \hat{k}_\mathrm{min}$ are used for testing the power-law hypothesis. Furthermore, we employ the symbol $k_{\mathrm{min}}$ to denote the smallest value of $k$ included in the analysis in cases where this value is not selected by the automatic procedure of the ML method (e.g., where it is chosen manually or where one sweeps through a range of values).

To find the optimal $\hat{k}_{\mathrm{min}}$, each unique value of $k$ present in the data is in turn used as $k_\mathrm{min}$, and a maximum likelihood estimate for the power-law exponent $\alpha$ is calculated considering only data points $k \geq k_{\mathrm{min}}$. Subsequently, the value of $k_{\mathrm{min}}$ minimizing the Kolmogorov-Smirnov (KS) distance between the cumulative distribution function (CDF) of the data points larger than or equal to $k_{\mathrm{min}}$ and the CDF of the fitted power-law model in the same region is selected as the optimal $\hat{k}_{\mathrm{min}}$. 

Next, the statistical plausibility of the best-fitting model is assessed with a goodness-of-fit test. Denoting the sample size with $n$ and the number of data points larger than or equal to $\hat{k}_{\mathrm{min}}$ with $n_{\mathrm{tail}}$, a number of synthetic datasets are generated with a semi-parametric bootstrap approach, where each data point is drawn from the best-fitting power-law model with probability $n_{\mathrm{tail}}/n$ and else from the empirical sample with $k < \hat{k}_{\mathrm{min}}$. A power-law model is then fitted to each of these bootstrapped samples, and the KS distance between the CDF of the original empirical distribution and its best-fitting power-law model is compared to the distribution of KS statistics between the generated synthetic datasets and their fitted models, and the test is considered to reject the power-law hypothesis if the fraction of KS statistics at least as extreme as the KS distance of the empirical distribution is smaller than a given $p$-value. Finally, the power-law model is compared to four alternative distributions (lognormal, exponential, Weibull, and truncated power law) using a normalized log-likelihood-ratio test originally presented by Vuong \emph{et al.}~\cite{vuong}. 

In Ref.~\cite{broido}, the sample is subsequently classified to one of six categories based on how strong evidence for the power-law hypothesis it is deemed to show. Here, we group these into two categories.  First, as in Ref.~\cite{broido}, a sample is said to show strong evidence for the power-law hypothesis if the following four conditions are met: 
\begin{enumerate}
\item The estimated exponent of the power law is between 2 and 3.
\item The number of data points in the tail, $n_{\mathrm{tail}}$, is at least 50. 
\item The goodness-of-fit test cannot reject the power-law hypothesis 
($p$-value $\geq0.1$). 
\item None of the alternative distributions is favored over power law in the log-likelihood-ratio test. 
\end{enumerate}
\noindent In Ref.~\cite{broido}, a sample falls into the category "not scale-free" if neither of conditions 3 or 4 is met. Consequently, we say that a sample shows some evidence for the power-law hypothesis if it fulfills at least one of these two conditions.

\begin{figure*}
\includegraphics[width=0.24\textwidth]{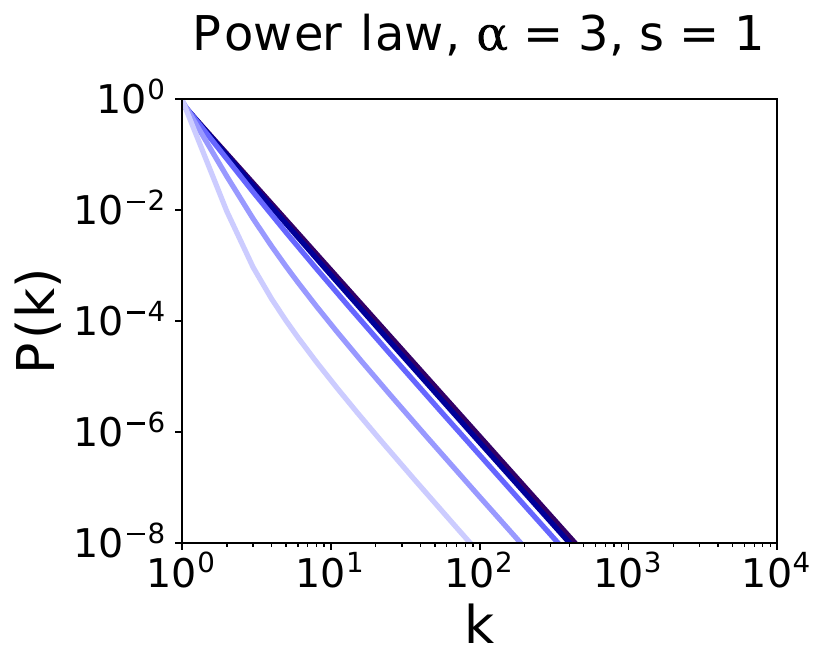}\includegraphics[width=0.24\textwidth]{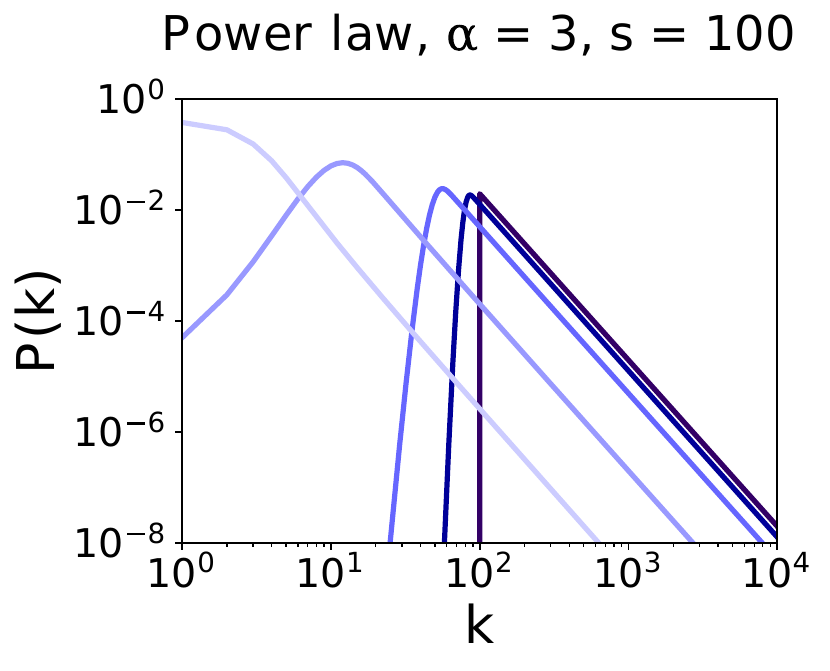}\includegraphics[width=0.24\textwidth]{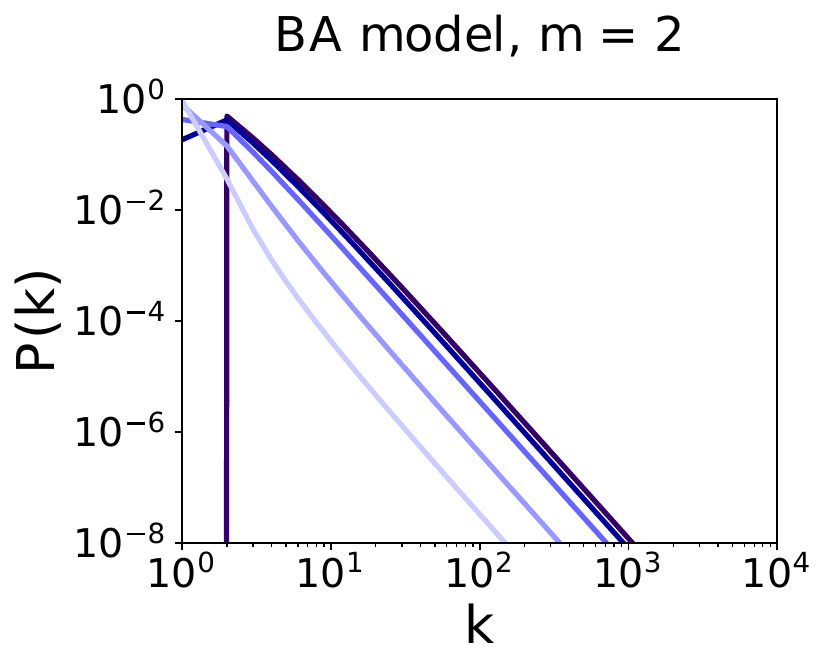}
\includegraphics[width=0.24\textwidth]{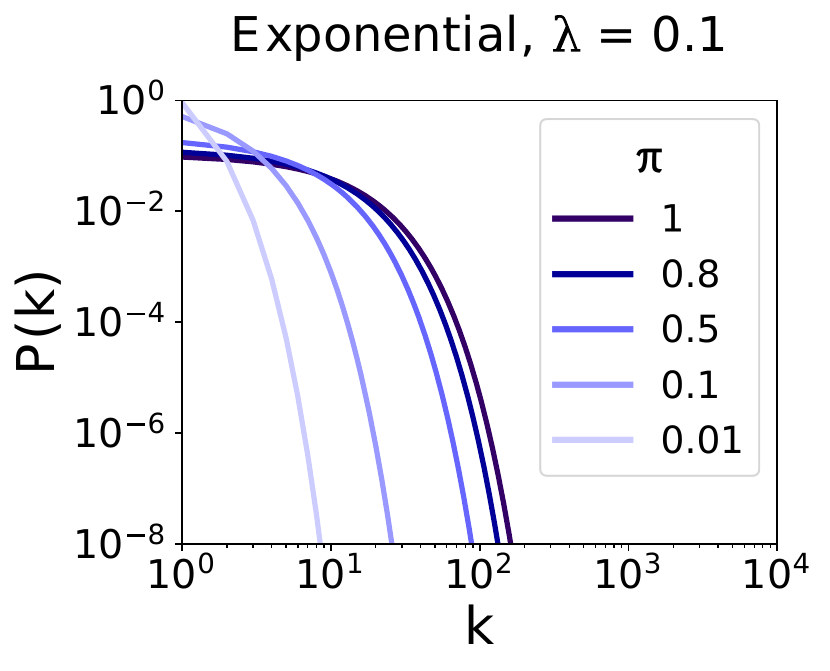}\\
\includegraphics[width=0.24\textwidth]{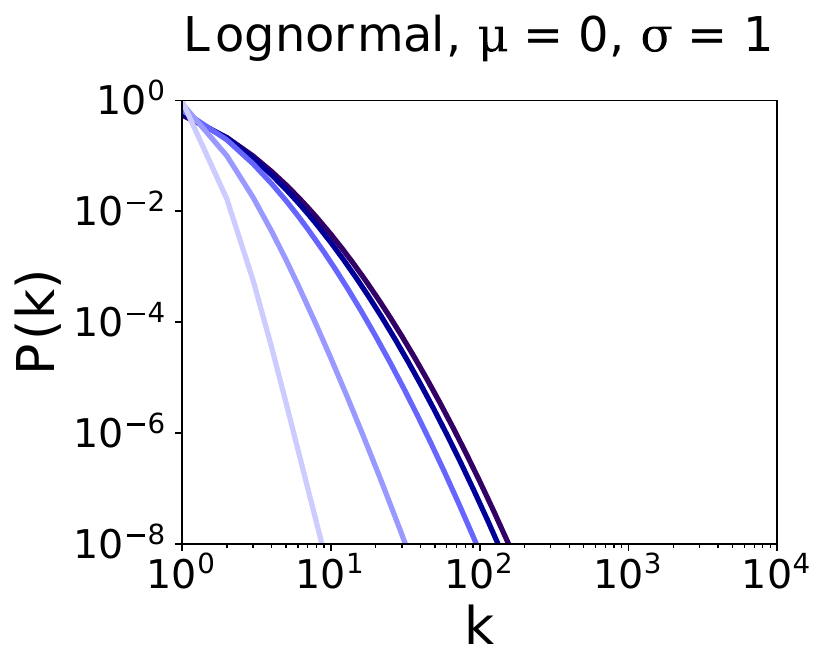}\includegraphics[width=0.24\textwidth]{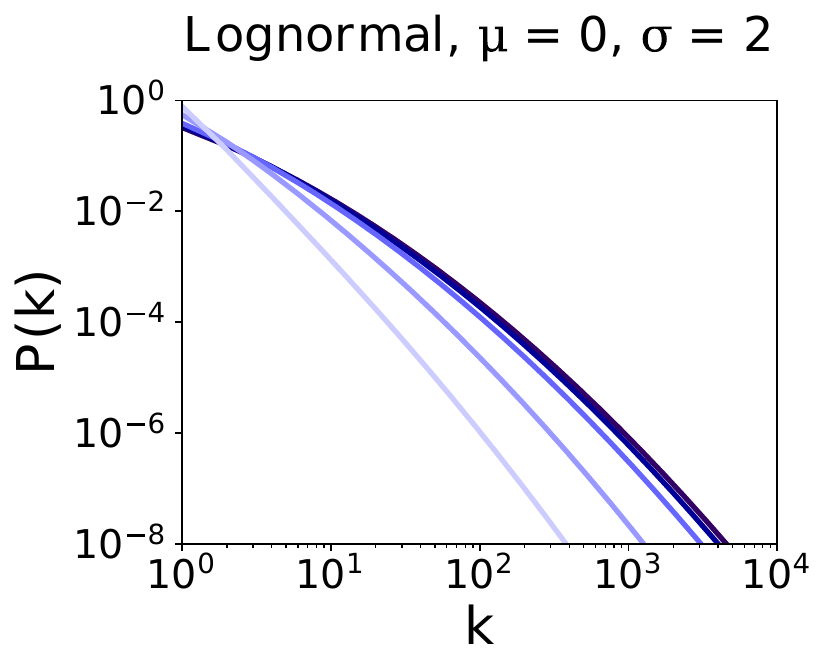}
\includegraphics[width=0.24\textwidth]{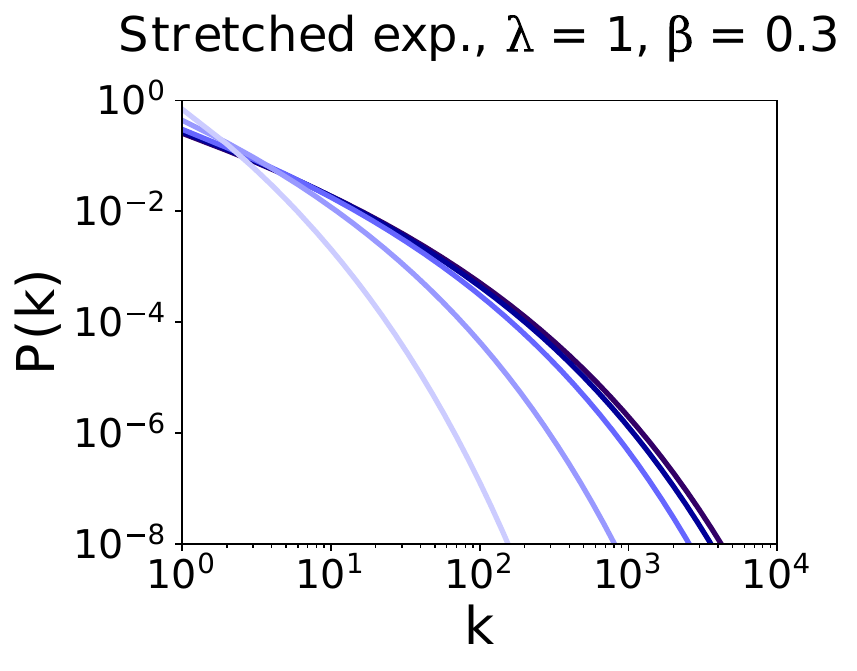}
\includegraphics[width=0.24\textwidth]{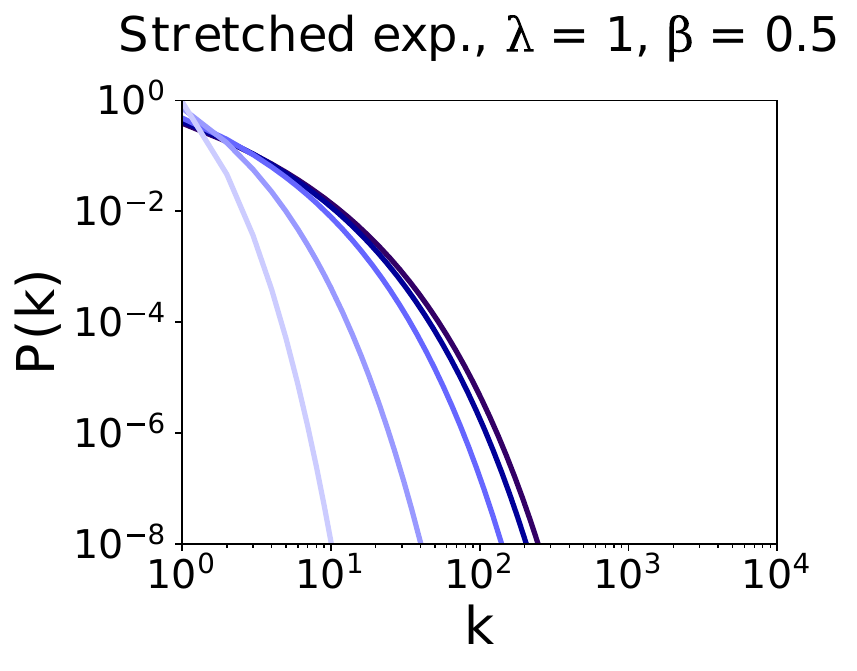}
\caption{\label{fig:sampled_pdfs} Examples of probability density functions (PDFs) of the subsampled distributions obtained by applying Eq.~(2) to the PDFs listed in Table 1 (procedure described in detail in section \ref{generating_analytical}). The beginning of the distribution's support before subsampling is denoted by $s$. Increasing $\mu$ and $\sigma$ of the lognormal distribution and decreasing $\lambda$ and $\beta$ of the stretched exponential distribution lead to heavier tails. Subsampling seems to increase the apparent degree of linearity on the log-log scale especially for the lognormal distributions.}
\end{figure*}

\subsection{\label{sec:estimators_voitalov}Extreme value (EV) method}

Voitalov \emph{et al.}~\cite{voitalov} broaden the definition of power-law distribution to encompass all regularly varying distributions. The complementary cumulative distribution function  (CCDF) of a regularly varying function is given by $\bar{F}(k) = l(k)k^{-(\alpha-1)}$, where $l(k)$ is a slowly varying function defined by the property $\lim_{k\to\infty} \frac{l(tk)}{l(k)} = 1$ for any $t>0$. All distributions with a probability density function given by $P(k) = l(k)k^{-\alpha}$ are regularly varying, but the converse is not true.

The extreme value (EV) method of Voitalov \emph{et al.}~is based on extreme value theory, which is concerned with the limit distribution of the sample maximum. Let $X_1, X_2, ..., X_n$ form a sample of independent and identically distributed random variables following a cumulative distribution function $F$. 
Then, $F$ is said to be in the maximum domain of attraction (MDA) of an extreme value distribution with tail index $\xi$, denoted by $F \in D_M(G_\xi)$, if there exist normalizing constants $a_n > 0$ and $b_n \in \mathbb{R}$ and a non-degenerate distribution function $G$ such that 
 \begin{align}
     &\lim_{n\to\infty} P \Big(\frac{\text{max}(X_1,X_2,...,X_n) - b_n}{a_n} \leq x \Big) \nonumber \\
     \
     =&\lim_{n\to\infty} [F(a_n x + b_n)]^n
     = G_\xi(x)\ \text{for all}\ x.
 \end{align}

\noindent By the Fisher-Tippett-Gnedenko theorem (see e.g.\ \cite{evt_theory}), there exist only three families of extreme value distributions $G_{\xi}(x)$ -- the Fr\'echet, Gumbel and reversed Weibull families -- each characterized by the tail index $\xi$ determining the shape of the distribution. Regularly varying distributions (both continuous and their discrete counterparts) form the MDA of the Fr\'echet distribution, for which $\xi > 0$. The power-law exponent of any regularly varying distribution in the MDA of the Fr\'echet distribution can be directly inferred from the tail index, 
\begin{equation}
    \alpha = \frac{1}{\xi} + 1.
\end{equation}
\noindent The reversed Weibull family, in turn, is characterized by a negative tail index, indicating that the distribution is bounded from above. Finally, for distributions belonging to the Gumbel MDA, the tail index equals zero.  The continuous lognormal and stretched exponential distributions belong to the MDA of the Gumbel distribution. As both lognormal and stretched exponential distributions belong to the class of long-tailed distributions \cite{nair}, also their discretized versions remain in the same MDA \cite{shimura}. In contrast, while the continuous exponential distribution is in the Gumbel MDA, its discrete counterpart does not belong to any MDA \cite{shimura}.

Voitalov \emph{et al.}~\cite{voitalov} propose estimating the power-law exponent $\alpha$ with three statistically consistent estimators of the tail index -- Hill, moments, and kernel -- which can be automated using a double bootstrap method for finding the optimal size of the tail considered in the estimation. The authors argue that due to the non-parametric nature of the regularly varying distribution family, it is impossible to quantify the probability that a given finite sample originates from a regularly varying distribution; however, if all three estimators return a clearly positive estimate of $\xi$, the observed sequence is likely to come from a regularly varying distribution. Consequently, Voitalov \emph{et al.}~classify a distribution as power law if all the considered estimators estimate the tail index to be over $1/4$. The limit is set to $1/4$ instead of zero to reduce the probability of falsely accepting the power-law hypothesis. The distribution is classified as not power law if at least one of the estimators returns a non-positive estimate. Else, the network is classified to be "hardly power law". In other words, the power-law exponent $\alpha$ needs to be between 1 and 5 for a sequence to be classified as a power law.

The first considered estimator, the Hill estimator \cite{Hill}, is statistically consistent for $\xi > 0$ and converges eventually to zero for $\xi = 0$. Given an ordered sample $X_{(1)} \geq X_{(2)} \geq ... \geq X_{(n)}$, the estimator operates on the $\kappa$ largest observations and gives an empirical estimate of the expected excess of the log-transformed distribution over threshold $\ln(X_{(\kappa+1)}$): 
\begin{equation}
    {\hat{\xi}}^{\mathrm{\mathrm{Hill}}}_{\kappa, n} = \frac{1}{\kappa}\sum_{i=1}^\kappa  \mbox{ln}\Big(\frac{X_{(i)}}{X_{(\kappa+1)}}\Big).
\end{equation}

\noindent As the other two estimators, the Hill estimator converges to the true tail index as $\kappa, n \to \infty$ and $ \kappa/n \to 0$.

The moments estimator \cite{dekkers} is an extension of the Hill estimator consistent for all $\xi \in \mathbb{R}$:
\begin{equation}
    \hat{\xi}^{\mathrm{Moments}}_{\kappa, n} = \hat{\xi}^{\mathrm{Hill}}_{\kappa, n} + 1 - \frac{1}{2}\Bigg( 1 - \frac{(\hat{\xi}^{\mathrm{Hill}}_{\kappa, n})^2}{\hat{\xi}^{\mathrm{Hill},2}_{\kappa, n}}\Bigg)^{-1}
    \label{eq:moments}
\end{equation}
\noindent where
\begin{equation}
    \hat{\xi}^{\mathrm{Hill},2}_{\kappa, n} = \frac{1}{\kappa}\sum_{i=1}^\kappa  \Big(\mbox{ln}\frac{X_{(i)}}{X_{(\kappa+1)}}\Big)^2.
\end{equation}
    
\noindent
For $\xi \geq 0$, $\lim\limits_{n\to\infty} \frac{(\hat{\xi}^{\mathrm{Hill}}_{\kappa, n})^2}{\hat{\xi}^{Hill,2}_{\kappa, n}} = \frac{1}{2}$ \cite{dekkers}, meaning that the moments estimator should converge to the value of the Hill estimator for distributions belonging to the MDA of either Fr\'echet or Gumbel families.

The kernel estimator \cite{groeneboom2003} is consistent for all $\xi \in \mathbb{R}$ as well.  The number of largest observations considered is determined by the bandwidth parameter $h$; approximately $nh$ observations are considered. The kernel estimator is given by 
\begin{align}
    \hat{\xi}^{\mathrm{Kernel}}_{h,n} &= \hat{\xi}^{\mathrm{pos}}_{h,n} - 1 + \frac{\hat{q}^{(2)}_{h,n}}{\hat{q}^{(1)}_{h,n}},\ \mbox{where}
    \label{eq:kernel_first}\\
    \ 
    \hat{\xi}^{\mathrm{pos}}_{h,n} &= \sum_{i=1}^{\lfloor nh \rfloor} \frac{i}{n} K_h \Big( \frac{i}{n} \Big) \mbox{ln} \Big( \frac{X_{(i)}}{X_{(i+1)}} \Big)  \label{eq14}\\
    &= - \int_0^h u K_h(u) \mbox{ }d \mbox{ ln} Q_n(1-u), \\
    \ 
    \hat{q}^{(1)}_{h,n} &= \sum_{i=1}^{\lfloor nh \rfloor} \Big(\frac{i}{n}\Big)^\gamma K_h \Big( \frac{i}{n} \Big) \mbox{ln} \Big( \frac{X_{(i)}}{X_{(i+1)}} \Big) \label{eq16}\\
    &= -\int_0^h u^\gamma K_h(u) \mbox{ }d \mbox{ ln} Q_n(1-u),\\
    \
    \hat{q}^{(2)}_{h,n} &= \sum_{i=1}^{\lfloor nh \rfloor} \frac{\partial}{\partial u} \Big[u^{\gamma +1} K_h(u) \Big]_{u=i/n} \mbox{ln} \Big( \frac{X_{(i)}}{X_{(i+1)}} \Big) \label{eq18}\\ 
    &= -\int_0^h \frac{d}{du} [u^{\gamma+1} K_h(u)] \mbox{ }d \mbox{ ln} Q_n(1-u),
    \label{eq:kernel_last}
\end{align} 

\noindent where 
the kernel $K_h(u)$ is given by $\frac{15}{8h} (1-(\frac{u}{h})^2)^2$, and $Q_n$ denotes the empirical quantile function  defined as $Q_n(u) = \inf \{x: F_n(x) \geq u \}$, where $F_n$ is the empirical distribution function.
Following \cite{voitalov}, we set the parameter $\gamma$ equal to $0.6$.
Note that we use $\lfloor nh \rfloor$ as the upper limit in the above summations while the upper limit is set to $\lfloor nh \rfloor -1$ in the code of Ref.~\cite{voitalov}.  As we illustrate in the Supplemental Material (SM) IB \cite{supplemental_material},  using either $\lfloor nh \rfloor$ or $\lfloor nh \rfloor-1$ as the upper limit yields identical results under some conditions; however, for a general $h$ (excluding $h=1$), the limit $\lfloor nh \rfloor$ yields correct results. We have modified the indexing in the code accordingly (see SM IB \cite{supplemental_material} for details).

\nocite{parzen}
\nocite{bartle}
\nocite{Bobkov_Ledoux_2019}

\section{\label{sec:simulating_subsampling}Simulating subsampling}

To investigate the performance of the estimators on subsampled distributions, we generate $n_0$ random numbers -- corresponding to the degrees of $n_0$ nodes -- from the desired distributions, subsample these simulated degree sequences using the previously described incident subgraph sampling strategy, and apply the ML and the EV methods to the obtained subsampled degree sequences. To avoid confusion, we use the term \emph{subsample} to refer to the samples obtained with incident subgraph sampling, while the term \emph{sample} refers to a sample from a given distribution to which the incident subgraph sampling has not yet been applied. Correspondingly, we denote the number of data points in the original sample by $n_0$ and use $n$ as general symbol for the sample size, whether of a subsample or the original sample.

\subsection{\label{sec:s}Simulation procedure}

We generate the samples with the exact search algorithm described in Ref.~\cite{clauset} and implemented in the Python package \emph{powerlaw} \cite{alstott}. In short, the algorithm operates by generating a random number $u \in (0,1]$ from a uniform distribution and returning the largest integer $i$ such that $\bar{F}(i) \geq u$, where $\bar{F}$ denotes the complementary cumulative distribution (CCDF) of the desired distribution. The probability density functions (PDFs) of the distributions are listed in Table  \ref{tab:table1}. After generating $n_0$ random degrees $k_j$ from a given distribution, we form a list where a unique identifier $j$ is repeated $k_j$ times for each node $j$, shuffle the list, and analyze the first $x$ identifiers of the list at a time, where $x = \mathrm{round}(\pi \sum_{j=1}^{n_0} k_j)$ and $\pi$ is the subsampling probability. With this method, the degree of each node $j$ in the subsample simply equals the number of identifiers $j$ in the selected list. We repeat the sampling procedure twenty times for each combination of parameters. Note that if two samples consist of the same number of data points, i.e.\ degrees, but the other sample comes from a distribution with a heavier tail, the sum of degrees is in general higher in that sample. Consequently, also the number of data points can be expected to be higher at any subsampling depth $\pi < 1$. Note also that the range of values of $k$ varies greatly for different distributions, becoming more restricted for distributions with smaller subsampling depths and lighter tails (such as the exponential distributions, stretched exponential distributions with $\beta=0.5$ and lognormal distributions with $\mu=1$). 

\begin{figure}
\centering
\includegraphics[width=0.7\textwidth]{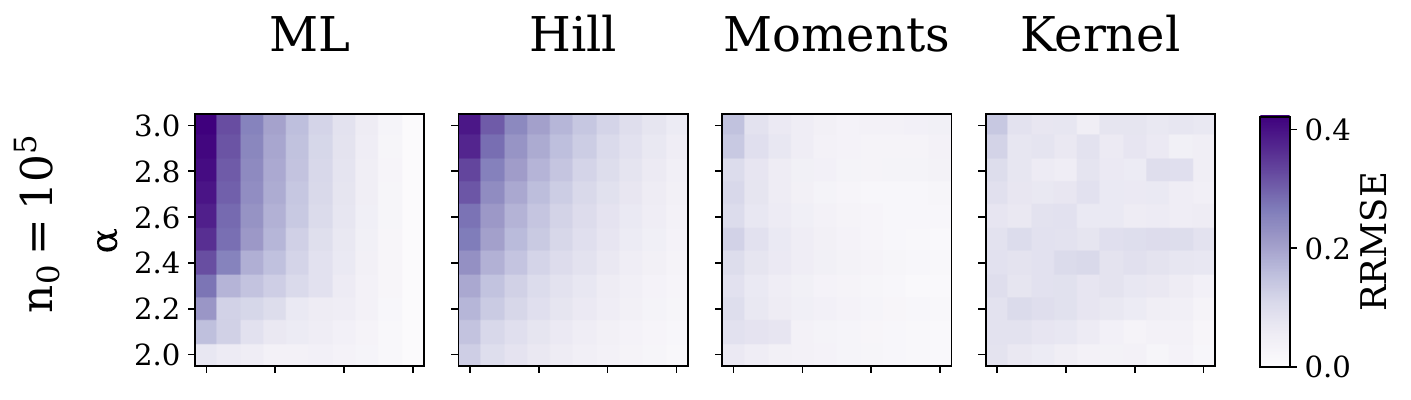}\\
\includegraphics[width=0.7\textwidth]{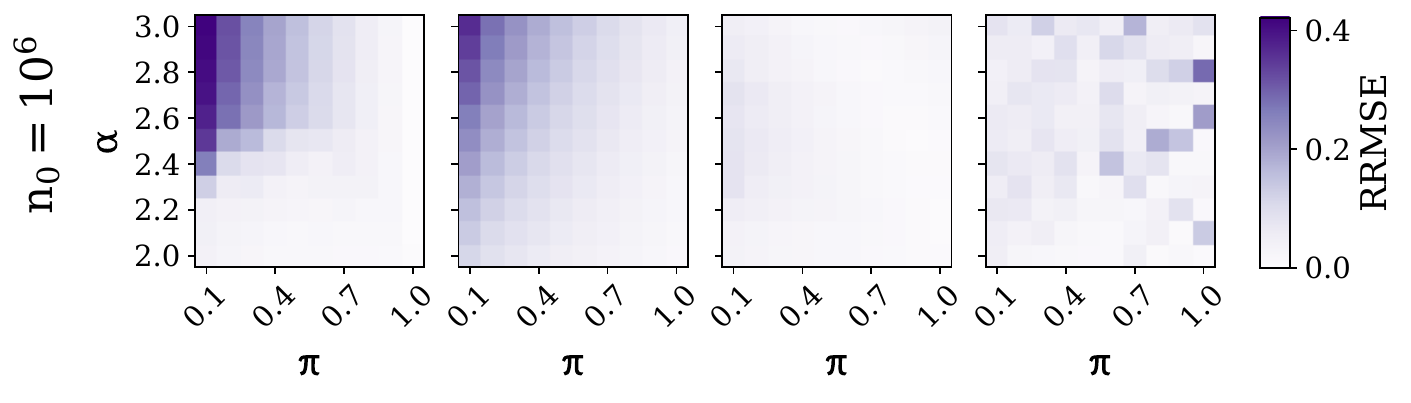}
\caption{\label{fig:wide}RRMSE values for the different estimates of the tail index $\xi$ as a function of the true power-law exponent $\alpha$ and the subsampling depth $\pi$. The darker the color, the greater the error in the tail index estimation. The columns correspond to the different estimators, while the subfigures in the same row share the same number of data points $n_0$ in the original sample. The number of (sub)samples used for calculating the RRMSE value in each cell is 20.}
\end{figure}

For the ML method, there are two available Python implementations of the original code in Matlab \cite{clauset}. We use the \emph{powerlaw} package by Alstott \emph{et al.}~\cite{alstott} for the model fitting as well as the log-likelihood-ratio tests. As the goodness-of-fit test is not implemented in this package, we use a modified version of the \emph{plpval} function from the implementation by Broido and Clauset \cite{broido}, where we have altered the function so that the model-fitting is again done with the \emph{powerlaw} package (see SM IA \cite{supplemental_material} for details). We have chosen to use a combination of these implementations due to an issue we encountered in the Broido and Clauset implementation regarding calculation of the normalization constant in cases where not all integers between the minimum and maximum values of the sample are present in the data (see SM I \cite{supplemental_material}). In the Alstott \emph{et al.}\ implementation, we have furthermore increased the number of allowed iterations when optimizing the parameters of alternative distributions with the \emph{scipy.optimize.fmin} function by adding the argument \emph{maxiter = 1000}. We do this to prevent termination of optimization before the parameters have converged (note that the variable \emph{warnflag} warning about failed convergence is not printed by default). 

For the EV method, we use the code released together with the article \cite{voitalov} with the modification discussed in Section IIB. We use the default parameters of the code, including adding a random uniform value $u \in [-0.5,0.5)$ to each discrete data point to enhance the performance of the estimators. A minor exception is the parameter $amseborder$, which 
reduces the likelihood of the double-bootstrap method choosing the order statistics from a region where the uniform noise dominates.  We use the default value $1.0$ in all (sub)samples where $k = 1$ is the smallest degree and set the value equal to the minimum degree for all other (sub)samples.

\begin{table}[!ht]
\caption{\label{tab:table1}
Probability density functions of the distributions, $P(k) = Cf(k)$. The distributions are normalized so that $\sum_{k=s}^\infty P(k) = 1$, where $s$ denotes the start of the distribution's support. The parameter ranges are the following: for power law, $\alpha > 1$; for discrete exponential (i.e.\ geometric), $\lambda > 0 $; for lognormal, $\mu \in (-\infty, \infty), \ \sigma > 0$ and for Weibull, $\lambda > 0, \ \beta > 0$ (for the heavy-tailed subclass that we refer to as stretched exp., $\beta \in (0,1)$). For the Barab\'asi-Albert (BA) model, $m$ is the number of nodes that each incoming node attaches to.}

\begin{tabular}{lll}
\toprule
\textrm{Distribution}&
\textrm{$C$}&
\textrm{$f(k)$}\\
\midrule
Power law  & $\zeta(\alpha,s)^{-1}$ & $k^{-\alpha}$\\
Exponential & $(1-e^{-\lambda})
e^{\lambda s}$ & $e^{-\lambda k}$ \\
Lognormal* & $\sqrt{\frac{2}{\pi \sigma^2}}\big[\text{erfc}(\frac{\ln s - \mu}{\sqrt 2 \sigma})\big]^{-1}$\ & $\frac{1}{k}\exp \big[-\frac{(\ln k - \mu)^2}{2\sigma^2}\big]$\\
Weibull* & $\beta \lambda e^{(\lambda s)^\beta}$  & $(\lambda k)^{\beta-1} e^{-(\lambda k)^\beta}$\\
BA model & & $\frac{2m(m+1)}{k(k+1)(k+2)} \text{ for } k \geq m$ \\
\bottomrule

\end{tabular}

\flushleft
* The discrete PDFs are approximated with the continuous PDFs listed here, as the distributions marked with a star lack an analytically defined discrete form. When generating random samples, the probabilities are obtained as $P(k) = F(k+0.5)-F(k-0.5)$, where $F(k)$ is the corresponding CDF, after which the probabilities are normalized by their sum.
\end{table}

\subsection{\label{sec:accuracy}Estimation accuracy of the power-law exponent}

As in Ref.~\cite{voitalov}, we assess the estimation accuracy of the power-law exponent with the relative root mean squared error (RRMSE). RRMSE is commonly used to measure the average error  in tail index estimation proportional to the real value of $\xi = \frac{1}{\alpha -1}$:
\begin{align*}
     &\mathrm{RRMSE} = \frac{\sqrt{\frac{1}{r} \sum_{j=1}^r (\hat{\xi}_j - \xi)^2}}{\xi},
\end{align*}
where $r$ is the number of samples. RRMSE is defined in terms of $\xi$ and not the power-law exponent $\alpha$ because $\alpha$ is defined only for distributions belonging to MDA of the Fr\'echet distribution, while $\xi$ characterizes all three extreme value distributions.

We observe that when $n_0$ -- i.e.\ the number of simulated data points in the original sample -- equals $10^5$, the moments and the kernel estimates stay fairly accurate even when the sequence is subsampled to a tenth of the original size [Fig.~\ref{fig:wide}, top row]. In contrast, both the maximum likelihood estimate of the ML method and the Hill estimate deviate further from the true power-law exponent $\alpha$ as the subsampling probability $\pi$ decreases and $\alpha$ grows larger (see SM II \cite{supplemental_material} for the values of $\hat{\alpha}$ and ${\hat{k}_{\mathrm{min}}}$ as well as the median sizes and maximum values of $k$ in the subsamples). Note that the number of data points for a certain subsampling depth differs greatly for different distributions, and the range of values $k$ gets more and more restricted for larger values of $\alpha$ and lower subsampling depths. However, increasing $n_0$ from $10^5$ to $10^6$ improves the accuracy only slightly [Fig.~\ref{fig:wide}, bottom row]. Due to the challenging shape of the heavily subsampled power-law sequences, all estimators tend to consider a too large part of the tail; for $\alpha$ close to three, the ML method consistently estimates $\hat{k}_{\mathrm{min}}$ to be one, which is clearly not an optimal choice considering the pronounced downward bending shape of subsampled power-law distributions (see the PDFs in Fig.~\ref{fig:sampled_pdfs}). Somewhat surprisingly, the kernel estimator becomes more unstable when $n_0 = 10^6$ and classifies some of the samples as belonging to the MDA of the Gumbel distribution. These misclassifications arise from a less-than-optimal estimation of the number of order statistics; almost all data points are taken into account, and consequently, the added random noise on the small degrees has a prominent effect on the estimate.  

The accuracy of the tail index estimation for small subsamples improves clearly if the original distribution has support only on higher values of $k$. In this case, the part of the distribution most affected by subsampling -- the smallest degrees -- has no probability mass to begin with and hence cannot dominate the estimation of $\hat{k}_{\mathrm{min}}$ or the number of order statistics. Consequently, the error of all estimators remains negligible for all distributions even for $\pi = 0.1$.

\subsection{\label{sec:level2}Classifying subsamples}

\noindent \textbf{Power-law distributions} 
\vspace{1mm}

\noindent
The EV method classifies all subsamples from power-law distributions correctly for all tested subsampling depths $\pi \in [0.1,1.0]$. As shown in the previous section, the moments and the kernel estimates stay fairly accurate even for small subsampling probabilities, and while the Hill estimator deviates from the true $\alpha$ with decreasing $\pi$, the estimates of $\alpha$ do not exceed the allowed upper limit for the tested subsampling probabilities. 

\begin{figure}[h!]
\centering
\includegraphics[width=0.6\textwidth]{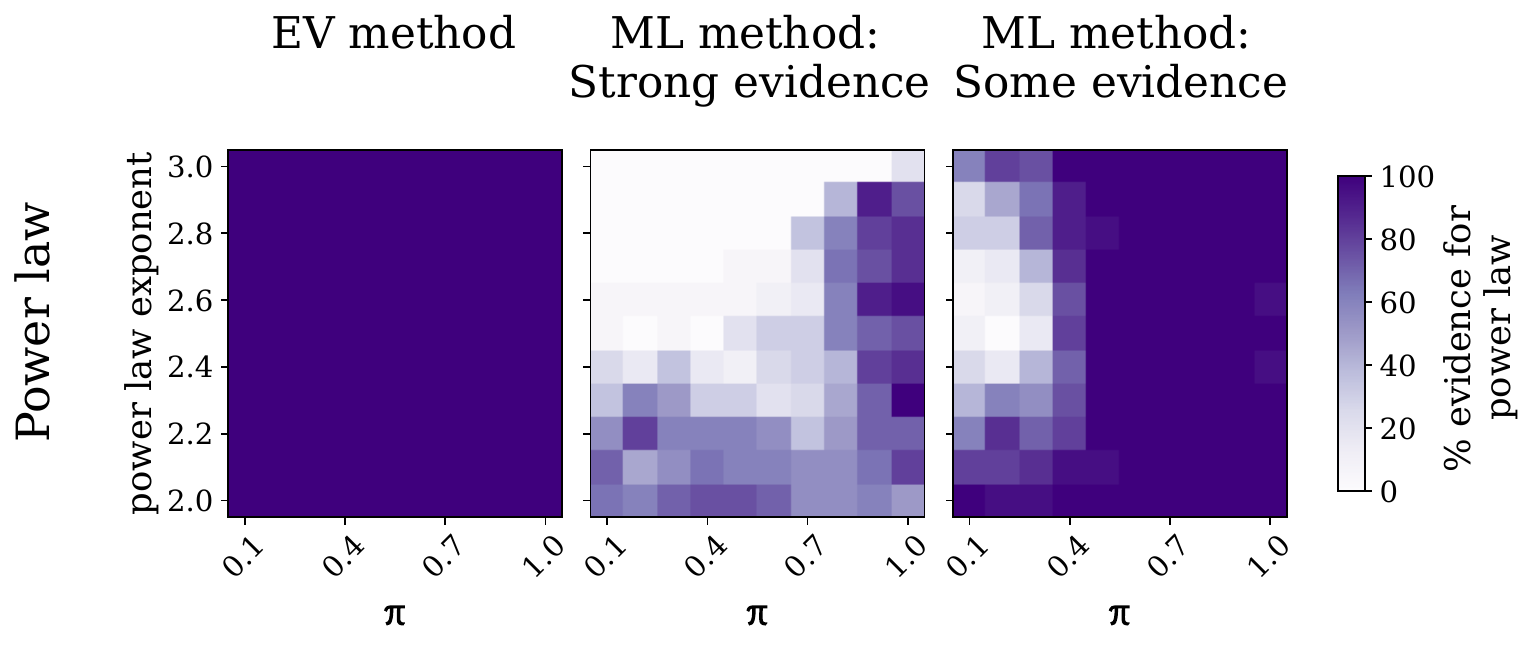}\\
\includegraphics[width=0.6\textwidth]{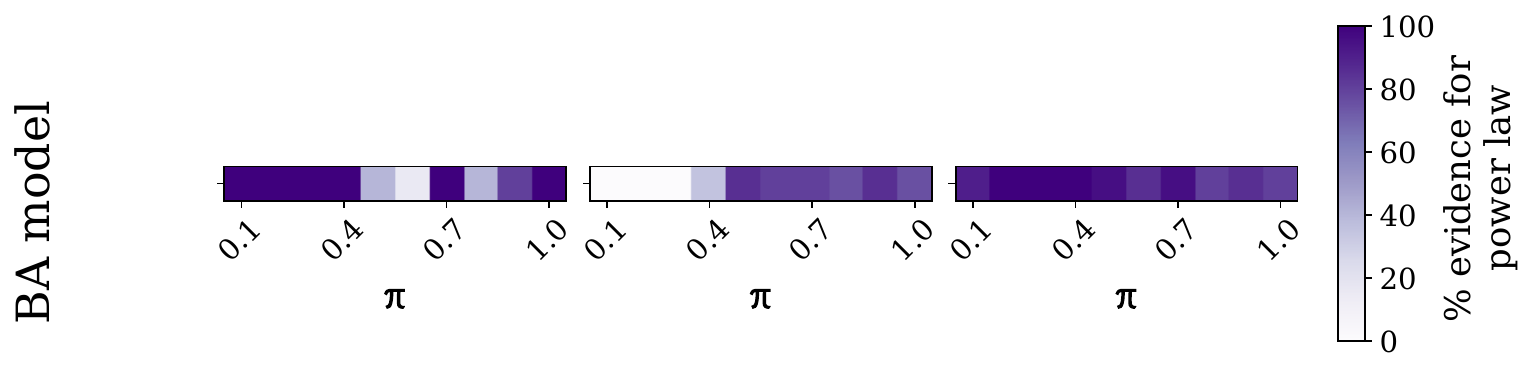}
\caption{Performance of the two methods with samples from power-law distributions as well as from degree distributions of networks grown by preferential attachment (BA model). The darker the color, the higher the fraction of subsamples correctly classified as power laws. The percentages are calculated over 20 samples with the number of data points before subsampling equal to $10^5$.}
\label{fig:powerlaw_identification}
\end{figure}

The ML method works less reliably on the subsamples [Figure \ref{fig:powerlaw_identification}]. The fraction of subsamples exhibiting strong evidence for the power-law hypothesis diminishes with decreasing $\pi$ and increasing $\alpha$, primarily due to the KS minimization driving the value of $\hat{k}_{\mathrm{min}}$ to a too low value, which results in an inaccurate estimate of $\alpha$ (see SM III \cite{supplemental_material} for performance of the individual criteria of the ML method and their different combinations). Consequently, the estimates of $\alpha$ often fall outside the allowed range $[2,3]$, and the goodness-of-fit test is more likely to reject the power-law hypothesis. This tendency of the ML method to select a too small $\hat{k}_{\mathrm{min}}$ is already noted in \cite{voitalov} and \cite{drees} with regard to some regularly varying distributions, and the problem is further aggravated by the probability mass concentrating  more heavily on small degrees as a result of subsampling. The log-likelihood-ratio test, however, continues to perform better than the goodness-of-fit test on the subsamples, resulting in a higher fraction of correctly classified subsamples when using the some evidence criteria.

In addition to pure power laws, we assess the behavior of the estimators with degree distributions of networks grown by preferential attachment. We simulate the Barab\'asi-Albert (BA) model, where  new nodes are added to the network one
by one and each new node forms $m=2$ connections to already existing nodes with probability $\frac{k_i}{\sum_j k_j}$, where $k_i$ denotes the degree of node $i$ and the index $j$ ranges over all the 
already existing nodes of the network \cite{barabasi}. The degree distribution of the network asymptotically approaches a power law with $\alpha = 3$ for $k \gg 1$. 

The strong evidence criteria of the ML method performs better with the subsamples from the BA model than with subsamples from a pure power law with $\alpha=3$. Performance is better because the fraction of nodes with degree one is smaller than for a pure power law, which leads to a more accurate estimation of $\hat{k}_{\mathrm{min}}$. However, subsamples with low subsampling depth are again falsely deemed to not to show strong evidence for the power-law hypothesis. 

The EV method, in turn, works less reliably on the subsamples from the BA model than on pure power laws [Fig.~\ref{fig:powerlaw_identification}]. This results from a tendency of the kernel estimator to occasionally estimate the subsample to belong to the MDA of the Gumbel distribution as a result of considering too large a fraction of the distribution's tail. 

\vspace{2mm}
\noindent \textbf{Exponential distributions} 

\begin{figure}[h!]
\centering
\includegraphics[width=0.5\textwidth]{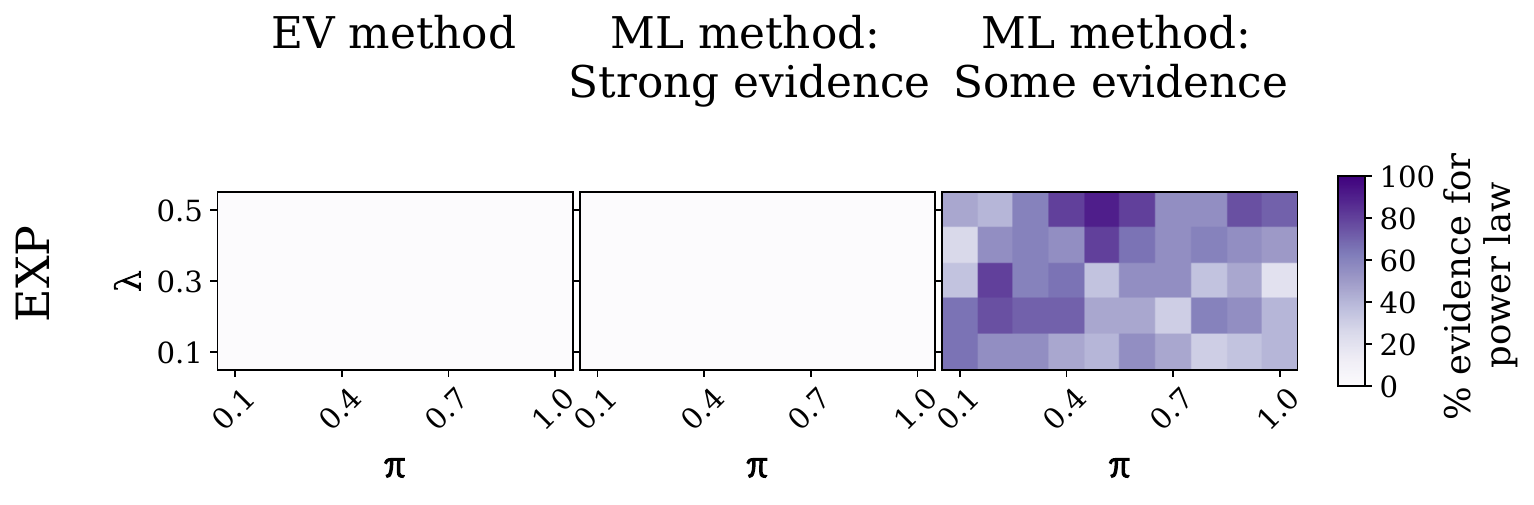}\\
\includegraphics[width=0.5\textwidth]{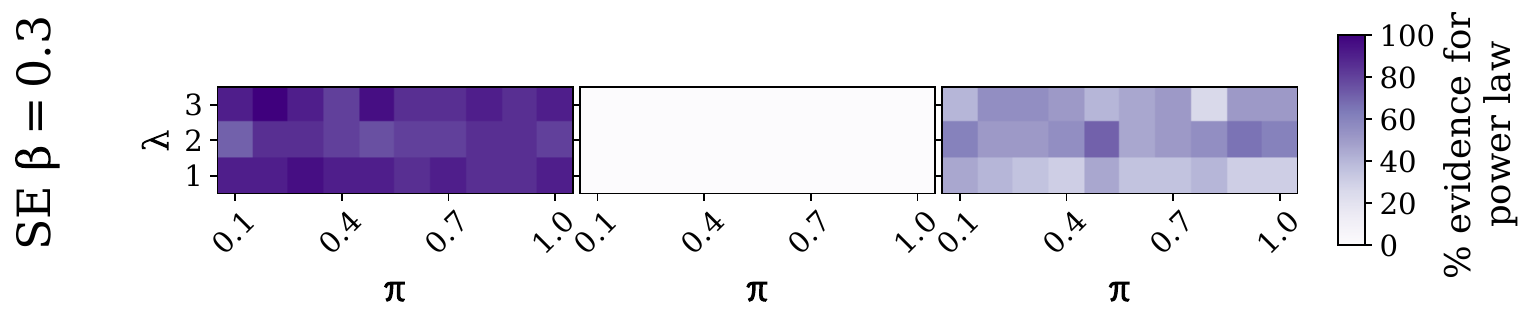}\\
\includegraphics[width=0.5\textwidth]{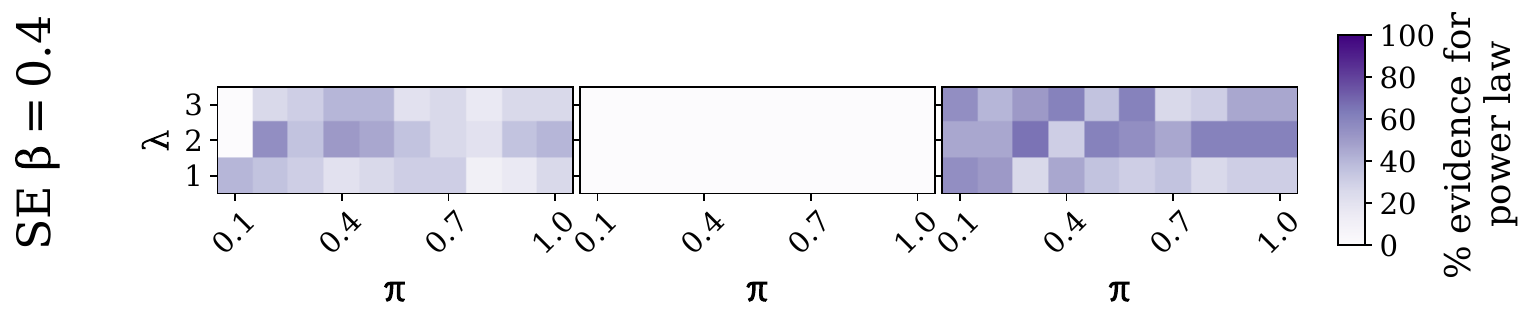}\\ 
\includegraphics[width=0.5\textwidth]{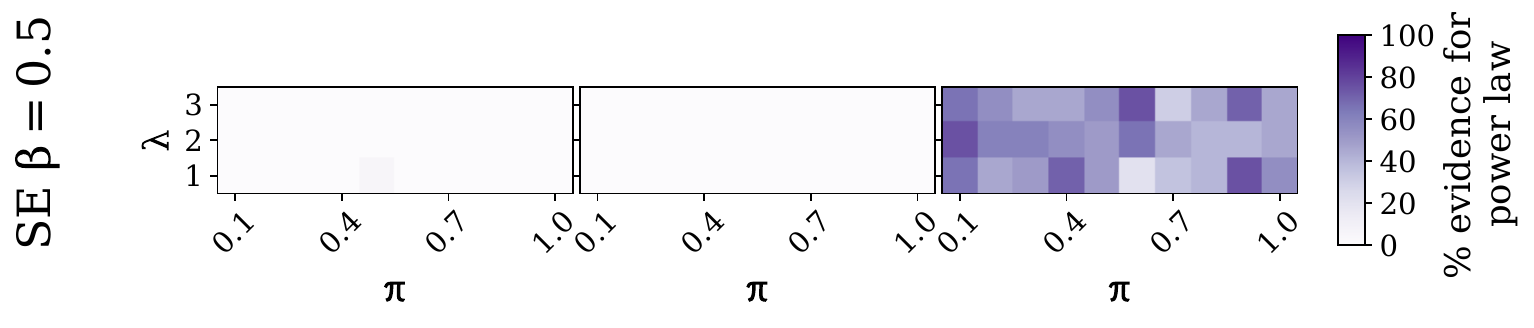}\\
\includegraphics[width=0.5\textwidth]{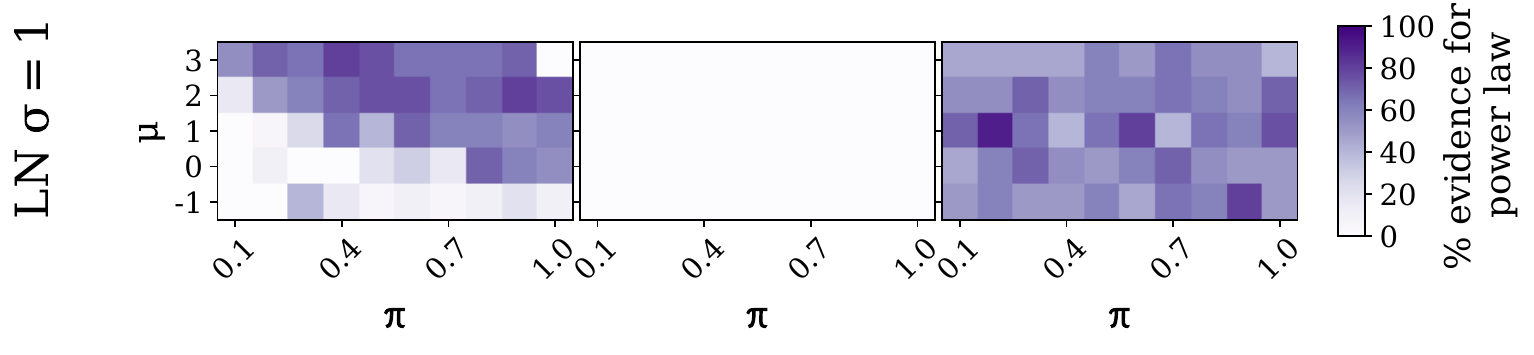}\\ 
\includegraphics[width=0.5\textwidth]{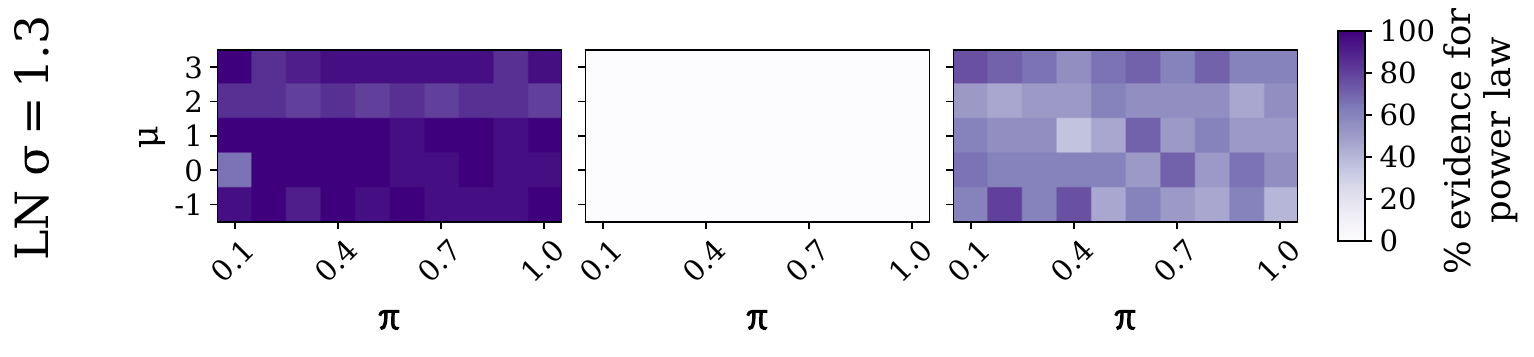}\\ 
\includegraphics[width=0.5\textwidth]{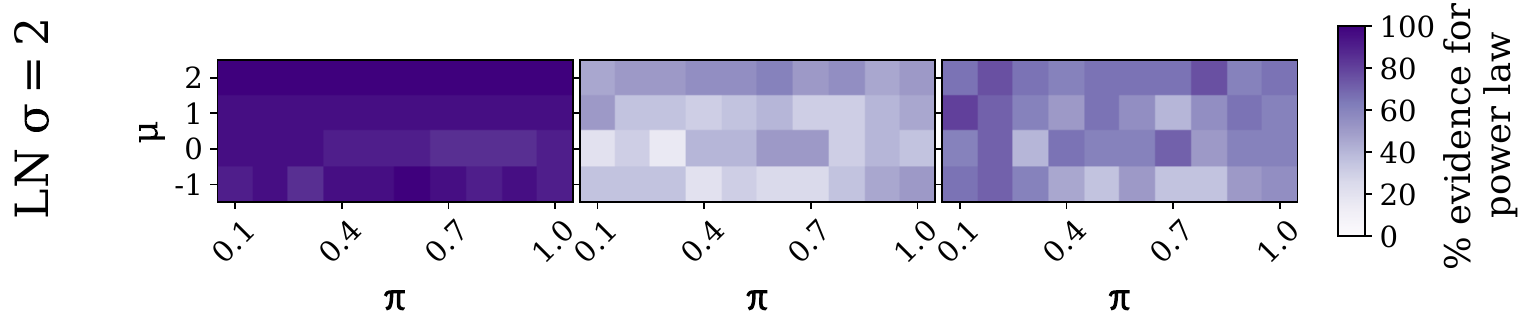}
\caption{Performance of the two methods with subsamples from exponential (EXP), lognormal (LN), and stretched exponential (SE) distributions. The darker the color, the higher the fraction of subsamples incorrectly classified as power laws. The number of data points in the original sample before subsampling is $10^5$ for all distributions except for the lognormal distribution with $\sigma = 2$, for which $n_0 = 10^4$ due to computational limitations. For the performance of the individual EV estimators, see SM III \cite{supplemental_material}. In general, combining the classifications of the three EV estimators outperforms any individual EV estimator on its own, supporting the use of the combined results to minimize false positive classifications.}
\label{fig:alternative_identification}
\end{figure}

\vspace{1mm}
\noindent All subsamples are classified correctly when using the strong evidence criteria of the ML method [Fig.~\ref{fig:alternative_identification}]. 
While the goodness-of-fit test and the log-likelihood ratio test do not always manage to reject the power-law hypothesis, the ML estimates of $\alpha$ stay consistently above the allowed upper limit, 3, which results in the lack of strong evidence.

The EV method correctly classifies all subsamples from exponential distribution with $\lambda \in[0.1,0.5]$  as non-power-law samples for all tested subsampling depths.

\vspace{2mm}
\noindent \textbf{Stretched exponential distributions}
\vspace{1mm}

\noindent According to the ML method, none of the subsamples with $\beta \in \{0.3,0.4, 0.5,0.7,0.9\}$ and $\lambda \in \{1,2,3\}$ show strong evidence for the power-law hypothesis [Fig.~\ref{fig:alternative_identification}]. A fraction of all (sub)samples shows some evidence for the PL hypothesis, but this fraction does not seem to be notably affected by the exact value of the subsampling probability $\pi$.

The EV method consistently classifies samples with $\beta \geq 0.5$ as non-power laws, but for $\beta = 0.4$ the rate of misclassification increases, and for $\beta = 0.3$ the majority of the original samples as well as the subsamples are classified as power laws. These incorrect classifications result from too slow convergence to the limiting extreme value distribution for small values of $\beta$, a problem noted with regards to $\beta=0.3$ already in \cite{malevergne2003}.

\newpage
\vspace{3mm}
\noindent \textbf{Lognormal distributions}
\vspace{1mm}

\noindent None of the lognormal sequences with $\sigma = 1$ exhibit strong evidence for the power-law hypothesis according to the ML criteria, again largely due to the estimates of $\alpha$ falling above the upper limit of the accepted range. These estimates decrease as $\sigma$ is increased and the tail becomes heavier, and consequently, some of the subsampled sequences with $\sigma=2$ are falsely deemed to show strong evidence.  

The EV method, in turn, erroneously classifies some of the samples with $\sigma=1$ as power laws, but -- somewhat surprisingly -- the rate of incorrect classifications seems to decrease with decreasing subsampling depth. This trend is more pronounced for distributions with smaller values of $\mu$. When $\sigma$ is increased to $1.3$, the vast majority of the original samples, as well as their subsamples, are falsely classified as power laws. This trend continues for $\sigma=2$; in addition, all three EV estimates tend to lie close to each other, which usually indicates that the classification should be reliable. 

As with the stretched exponential distribution, the unreliability of the EV method for lognormal distributions with larger values of $\sigma$ originates from too slow convergence to the asymptotic EV distribution. This dramatic slowing down as $\sigma$ increases can be seen in the convergence rate in the von Mises condition on which the consistency analysis of the kernel method is based on (Eq.~1.1 in \cite{groeneboom2003}),
\begin{equation}
    \lim_{k_{\mathrm{min}} \to \infty }\!M(k_{\mathrm{min}})    \!=\! \xi, \ \mathrm{ where} \ M(k_{\mathrm{min}}) \!=\!\frac{d}{d  k_{\mathrm{min}}} \frac{1\! -\!F(k_{\mathrm{min}})}{F'(k_{\mathrm{min}})}.
\end{equation}

\noindent For the continuous lognormal distribution with $\mu=2$ and $\sigma=2$, we observe that the value of $M(k_{\mathrm{min}})$ falls for the first time under the threshold 1/4 (used by the EV method to separate power laws from non-power-laws) when $k_{\mathrm{min}} > 17 \ 000 \ 000$. In contrast, when the parameter $\sigma$ is decreased to $1$, the value of $M(k_{\mathrm{min}})$ falls under the threshold already around  $k_{\mathrm{min}} \approx 73$. 

\vspace{4mm}

\noindent In conclusion, it seems that the EV method and the strong evidence criteria of the ML method suffer from opposite tendencies; while the strong criteria  tends to produce false negatives, the EV method is prone to falsely accepting the power-law hypothesis for some of the tested alternative heavy-tailed distributions. The some evidence criteria of the ML method suffers from a substantial rate of false positives as well. Overall, analyzing the performance of the individual criteria of the ML method shows that the log-likelihood ratio criteria performs better than the goodness-of-fit test both with regard to the rate of false positives and false negatives, and hence its overall performance is better than that of the \emph{some evidence} criteria (see SM III \cite{supplemental_material}). Whether one prefers to use this criterion alone, the strong evidence criteria or the strong evidence criteria without the goodness-of-fit test ultimately depends on how much emphasis one places on avoiding false positives at the expense of an increased false negative rate.

Interestingly, it would seem that the methods' ability to reliably reject the power-law hypothesis for subsamples from alternative distributions depends to a great extent on their ability to classify the original sample correctly. As an exception to this trend we observed that the accuracy of the EV method seemed to increase for some lognormal distributions as a result of subsampling. However, as the results of the simulations depend rather heavily on the sample size $n$, it is not clear whether this exception originates from the properties of the subsampled distributions or simply from the smaller amount of data points in the substantially downsampled data. In addition, trends visible with either larger or smaller sample sizes might go unnoticed in the simulations. To resolve these questions, we now turn to examine the behavior of the estimators on the theoretical subsampled probability distributions.

\section{Performance on theoretical subsampled probability distributions of the heavy-tailed alternatives} \label{analytical_results}

In this section, we analyze the theoretical subsampled probability distributions of the heavy-tailed alternatives and assess whether correctly classifying these lognormal and stretched exponential distributions as non-power-law becomes more challenging as the subsampling depth decreases. We put all considerations of the sample size $n$ aside and assess the behavior of the estimators on the distributions in the limit of large $n$ (derivations in SM IV \cite{supplemental_material}), under the assumption that the subsampling depth or the choice of the smallest considered degree $k_{\mathrm{min}}$ does not significantly affect the rate at which an estimator converges towards its limiting value. Note that we do not consider here how the size of the tail to be considered is in practice estimated for empirical or simulated data with either the double bootstrap procedure of the EV method or the KS minimization of the ML method, since this selection depends on the sample size $n$. Instead, we manually vary the value of $k_{\mathrm{min}}$ and assess how the analytically calculated estimates change as a result.

\subsection{Generating theoretical distributions}
\label{generating_analytical}

Given a discrete probability density function $P(k)$ (listed in Table \ref{tab:table1}), the theoretical subsampled distributions are obtained by brute force using Equation \ref{eq:subsampling} with $i = 10^5$ as the upper limit in the summation. Some examples of the theoretical subsampled distributions are shown in Fig.~\ref{fig:sampled_pdfs}. As the upper limit is finite, these numerically obtained subsampled distributions are not exact.
To exclude the possibility of numerical imprecision confounding our results, we verify that increasing the upper limit to $i=10^6$ does not visibly change the results for the largest and the smallest subsampling depth in Figs.~\ref{fig:edge} and \ref{fig:analytical_clauset} (as well as Fig.~9 in the SM \cite{supplemental_material}) for the parameter combinations producing the heaviest tails. We refrain from considering parameter combinations requiring an even larger limit, as already the limit $i=10^6$ is computationally very heavy. 

Since lognormal and stretched exponential distributions do not have an analytically defined discrete form, we approximate the original discrete distributions before subsampling by calculating the point-wise probabilities $P(k) = f(k)$ (the notation corresponds to that in Table \ref{tab:table1}) from $k=1$ up to $k=10^5$ and normalizing the probabilities by their sum. To verify that our conclusions are not affected by the chosen discretization strategy, we repeat the analysis for one representative distribution from both lognormal and stretched exponential families with two other discretization methods; one, where $P(k) = F(k+1) - F(k)$, and another, where $P(k) = F(k+0.5) - F(k-0.5)$. While the exact numerical limit values of the estimators for a given $k_{\mathrm{min}}$ depend on the discretization strategy, the conclusions remain unaffected.

When analyzing the behavior of the EV estimators, we treat the probability distribution functions not as discrete functions but as step functions, where the probability remains constant from $k-0.5$ up to $k+0.5$ for each $k$. We do this to mirror the empirical analysis as closely as possible. This transformation corresponds to adding random uniform noise to each data point, as is done with the simulated data.

\subsection{Extreme value (EV) method} \label{analytical_voitalov_results}

As previously discussed, the successfulness of the EV method depends to a great extent on the distribution's rate of convergence to the asymptotic extreme value distribution; correct classification of the distribution's type becomes more difficult if this rate slows down considerably. In practice, this would mean that the fraction of the distribution's tail that can be used to correctly identify the maximum domain of attraction  (and hence the type of the distribution) becomes too small, meaning that the sample size $n$ would need to be unreasonably large to capture this part of the tail. To assess the effect of subsampling on the estimators' performance, we examine how the fraction of probability mass in the distribution's tail that allows for correct identification of the distribution's type changes as the subsampling depth decreases. In the following, we refer to this fraction as the usable part of the tail. In the case of lognormal and stretched exponential distributions, this fraction corresponds to the value of the subsampled distribution's CCDF at the smallest value of $k_{\mathrm{min}}$ (not necessarily an integer in this context) for which the tail index falls below the limit of $1/4$ allowing the distribution to be correctly classified as non-power-law according to the criteria of the EV method. We denote this value of $k_{\mathrm{min}}$ with $k^\ast$ and assess it with an accuracy of 0.01. 

\begin{figure}[h!]
\centering
\includegraphics[width=0.6\textwidth]{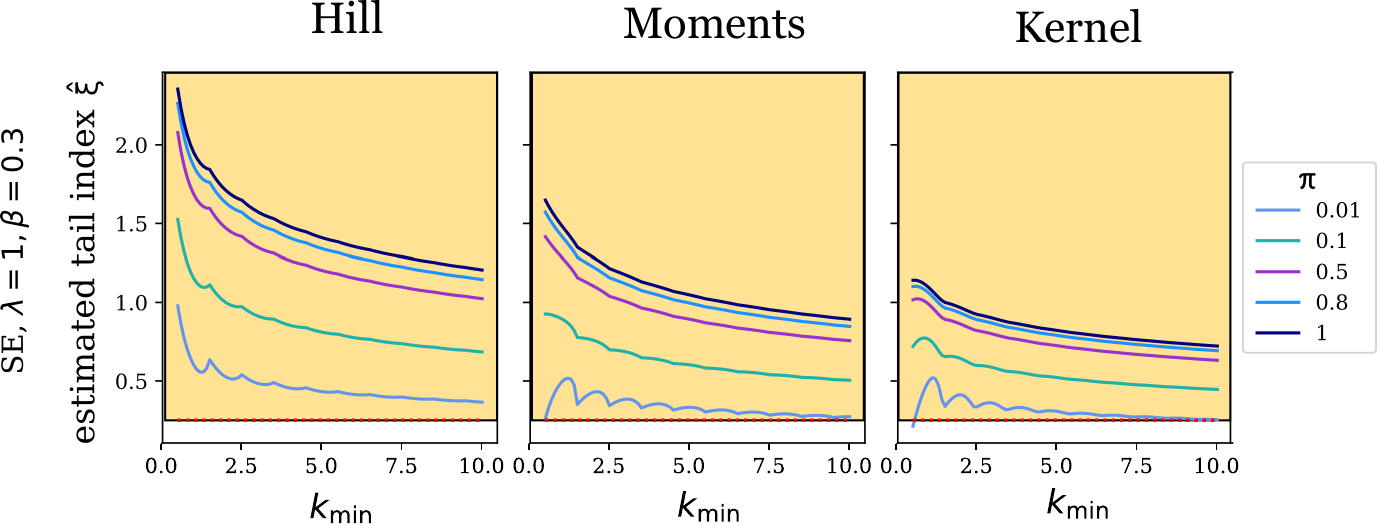} \\
\includegraphics[width=0.6\textwidth]{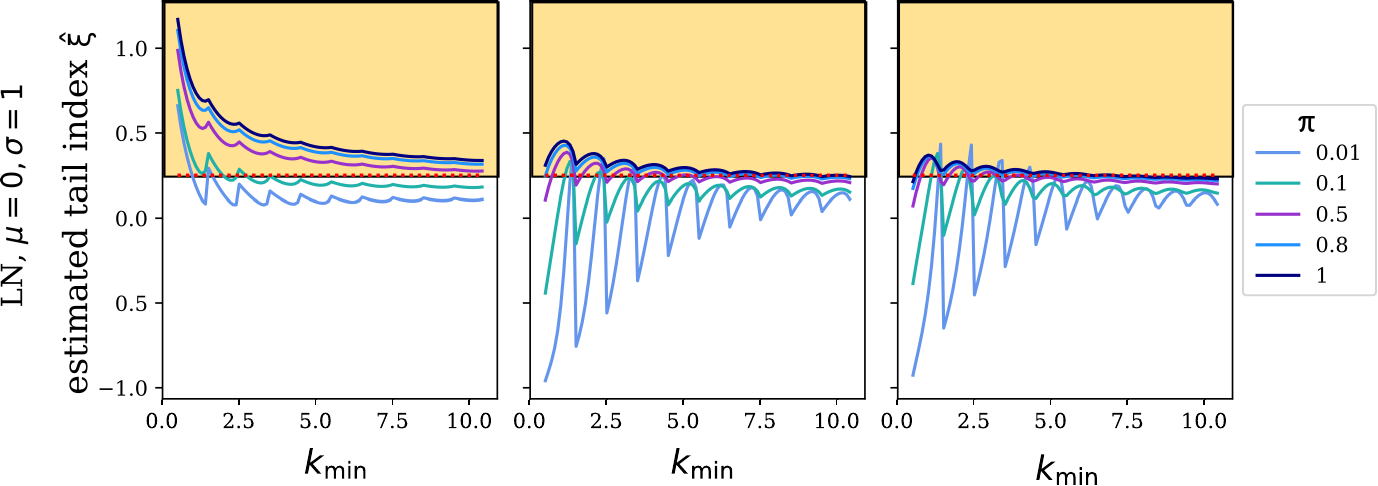}
\caption{Examples of the behavior of the estimated tail index $\hat{\xi}$ in the limit of large $n$ as a function of the smallest considered degree $k_{\mathrm{min}}$ for a stretched exponential distribution (SE, top row) and a lognormal distribution (LN, bottom row) with different subsampling depths $\pi$. The yellow shaded area ($\hat{\xi} > 1/4$) marks the values of $\hat{\xi}$ for which the distribution is incorrectly classified as a power law.}
\label{fig:analytical_voitalov}
\end{figure}

\begin{figure}[h!]
\centering
\includegraphics[width=0.5\textwidth]{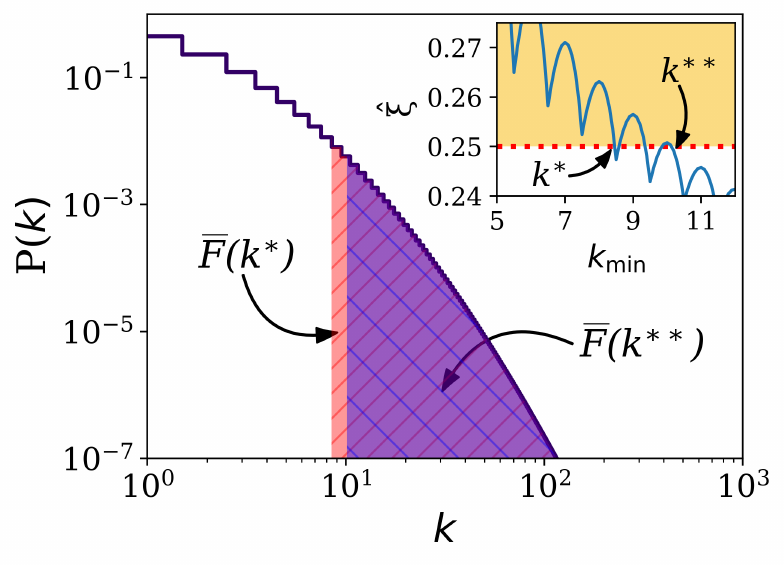}
\caption{Illustration of the fractions $\bar{F}(k^{\ast})$ and $\bar{F}(k^{\ast \ast})$ for the lognormal distribution with $\mu=1$, $\sigma=1$ and subsampling depth $\pi=0.4$. The inset shows the analytical limit value of the moments estimator as a function of the smallest considered degree $k_{\mathrm{min}}$. At $k_{\mathrm{min}}=k^{\ast}$, this limit value falls for the first time (and at $k_{\mathrm{min}}=k^{\ast \ast}$, permanently) under the threshold 1/4 allowing for correctly classifying the distribution as non-power-law. The shaded areas $\bar{F}(k^{\ast})$ and $\bar{F}(k^{\ast \ast})$ display the corresponding values of the theoretical subsampled probability distribution's CCDF, i.e.\ the fractions of the distribution's tail that can be used to correctly identify the distribution's type according to the criteria of the EV method.}
\label{fig:illustrative}
\end{figure}

As we are operating with step functions, the estimates do not necessarily decrease monotonically as a function of the smallest considered degree $k_{\mathrm{min}}$ (see Fig.~\ref{fig:analytical_voitalov} for examples), and the estimate might consequently rise above the threshold $1/4$ even for $k_{\mathrm{min}} > k^\ast$. 
Consequently, it may be more informative to assess the CCDF at the point where the estimate falls permanently under the threshold of $1/4$. We denote this point with $k^{\ast \ast}$ and the corresponding value of the CCDF with $\bar{F}(k^{\ast \ast})$ (see Fig.~\ref{fig:illustrative} for illustration). In the following, we refer to $\bar{F}(k^{\ast \ast})$ as the \emph{fully usable} fraction of the tail and to $\bar{F}(k^{\ast})$ as the \emph{potentially usable} fraction. For some distributions the values  $k^{\ast}$ and $k^{\ast \ast}$ coincide. In general, the moments and the kernel estimators suffer from oscillations more than the Hill estimator, and the oscillations grow larger the more heavily the probability mass concentrates on the smallest degrees. This happens when subsampling depth decreases or parameters of the distributions are changed to produce a lighter tail. 

\begin{figure*}[htp]
\centering
    \subfloat[Lognormal with $\sigma=1$]{\label{fig:ln_sigma1}\includegraphics[width=.5\linewidth]{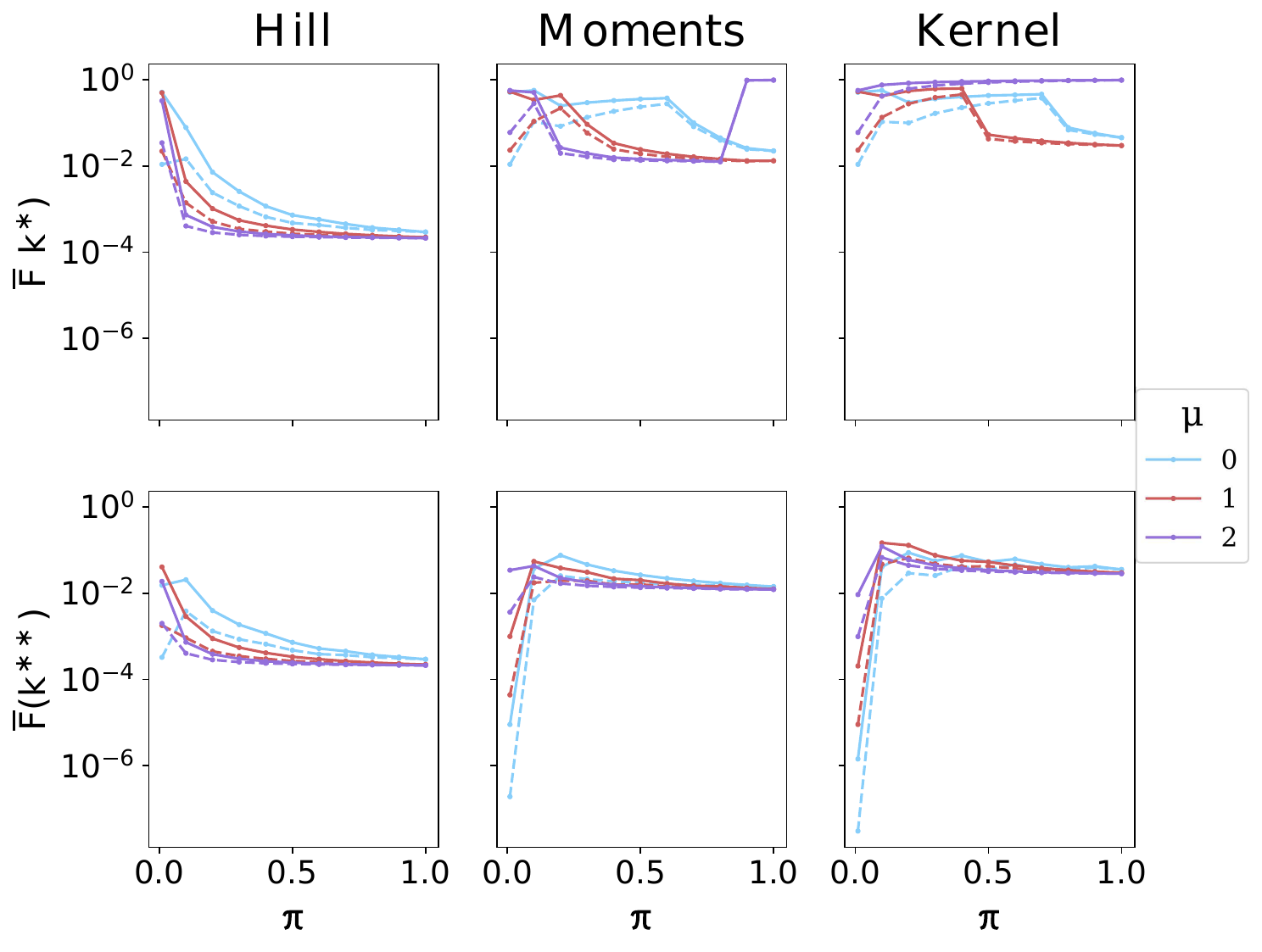}}
    \subfloat[\label{fig:beta03}Stretched exponential with $\beta=0.3$]{\includegraphics[width=.5\linewidth]{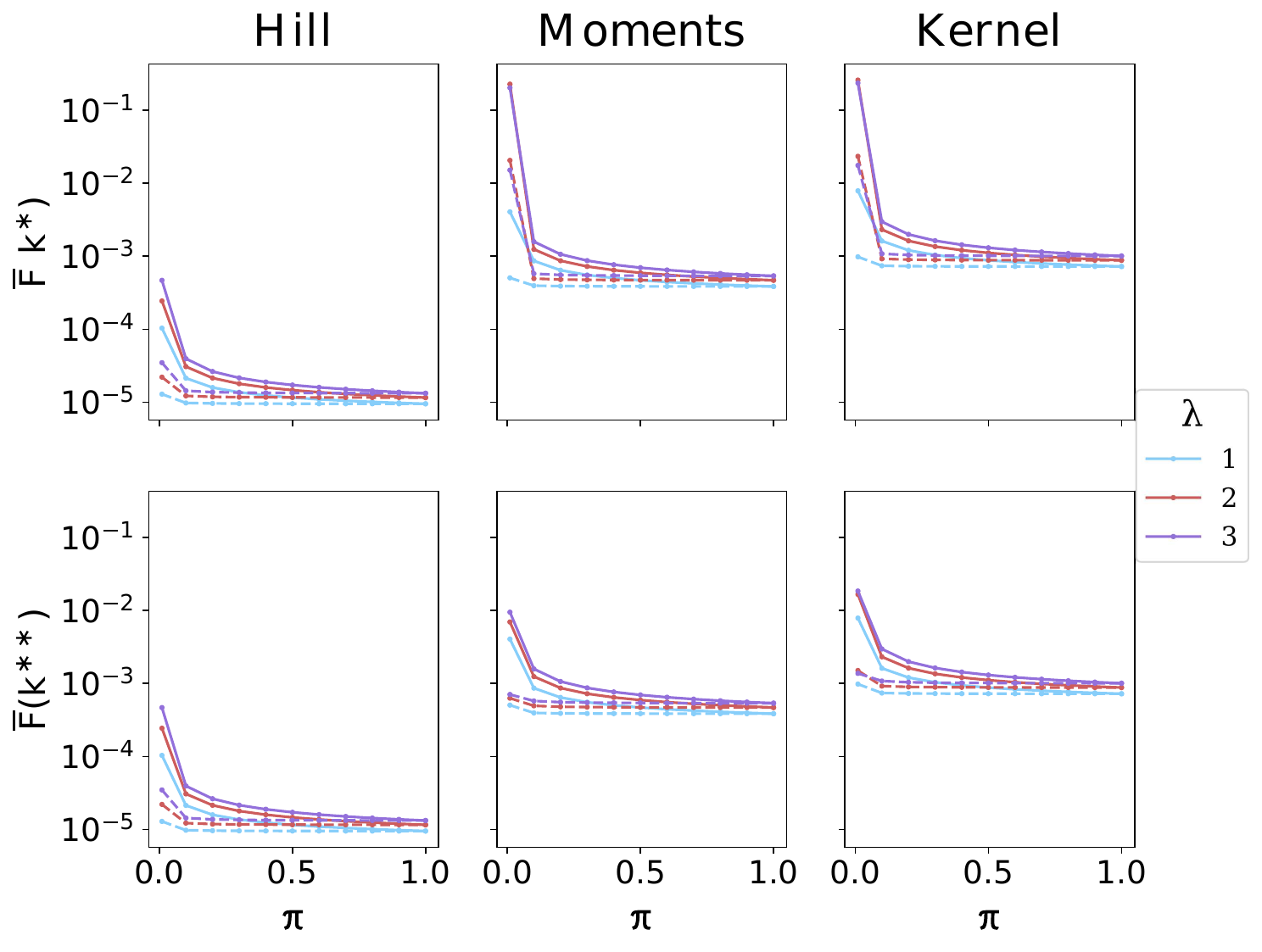}}
  \caption{Value of the CCDF at the smallest value of $k_{\mathrm{min}}$ (smallest degree considered in the analysis) at which the limit value of the EV estimators falls for the first time (top row) and permanently (bottom row) below the threshold 1/4 resulting in the distribution being correctly classified as non-power-law. Larger values of the CCDF indicate that a larger fraction of the distribution's tail allows for correctly identifying the distribution. The value of the CCDF is expressed both relative to the subsampled distribution (solid lines) and to the original distribution (dashed lines). 
  In the top row of subfigure a), the abrupt fall of $\bar{F}(k^*)$ of the moments estimator for the lognormal distribution with $\mu = 2$ at $\pi=0.8$ originates from the fact that we start to assess the estimators performance at $k_{\mathrm{min}} = 1.0$ instead of $k_{\mathrm{min}} = 0.5$ mimicking the default parameters of the code. For $\mu = 2$, the estimates for all tested subsampling probabilities fall originally below 1/4 at $k_{\mathrm{min}} = 0.5$ and subsequently rise above this threshold, but for $\pi=1.0$ this rising happens more slowly and the moments estimate still lies below 1/4 at $k_{\mathrm{min}} = 1$.}
  \label{fig:edge}
\end{figure*}

When comparing the usable fraction of the tail for different subsampling depths, we are essentially asking how the expected number of nodes from the usable part would change if we were to randomly draw samples of equal size from distributions of different subsampling depths. To understand how the expected number of nodes changes in cases where the sample sizes are unequal as a result of subsampling, we assess the fraction of the tail allowing for correct identification relative to the original distribution ($\pi=1$) as well. We do this by multiplying the obtained value of $\bar{F}(k^\ast)$ or $\bar{F}(k^{\ast \ast})$ with the fraction of the total non-normalized probability mass of the subsampled distribution, i.e., the probability $1 - P_s(k = 0)$, where $P_s(k=0)$ is obtained using Eq.~\ref{eq:subsampling}.

Interestingly, the lognormal and the stretched exponential distributions tend to become easier to separate from power laws when subsampled. Identification becomes easier in the sense that the fully usable fraction
$\bar{F}(k^{\ast \ast})$ of the tail tends to slightly increase as the subsampling depth decreases. However, this effect is far from linear, and if the tail is not heavy enough, $\bar{F}(k^{\ast \ast})$ can drop  dramatically for very small values of $\pi$ due to the violently oscillating pattern of the estimator [Fig.~\ref{fig:ln_sigma1}]. 

For the Hill estimator also the potentially usable fraction $\bar{F}(k^{\ast})$ tends to become larger for smaller subsampling probabilities, even to the extent that this fraction of probability mass allowing for correct classification increases when expressed relative to the original distribution before subsampling. Since the oscillations are more pronounced for the moments and the kernel estimators, the potentially usable fraction behaves in a more unpredictable manner for these estimators. However, if the tail of the distribution is heavy enough, this fraction seems to increase consistently even for these estimators [Fig.~\ref{fig:beta03}]. Overall, these results imply that the previous finding of lognormal subsamples being more accurately classified for lower subsampling depths may result from better discriminability of the subsampled distribution and not only on the smaller sample size.

\subsection{Maximum likelihood (ML) method} \label{analytical_clauset_results}
As in the previous section, we examine how subsampling affects the ease of classification by varying the smallest considered degree $k_{\mathrm{min}}$ and calculating the corresponding limit values of the estimators as $n \to \infty$ for the theoretical subsampled distributions. To allow comparison between different subsampling depths, we display the results not as a function of $k_{\mathrm{min}}$ but as a function of the value of the CCDF $\bar{F}(k_{\mathrm{min}})$. In contrast to the EV method, the ML method classifies empirical data using the $p$-values associated with the estimates, and hence we cannot directly assess any threshold
for $\bar{F}(k_{\mathrm{min}})$ above which the power-law hypothesis would be rejected. However, we can analyze whether the criteria for the power-law hypothesis would be more likely to be met in comparison with a different subsampling depth for a given $\bar{F}(k_{\mathrm{min}})$. 

\begin{figure*}
\includegraphics[width=1\textwidth]{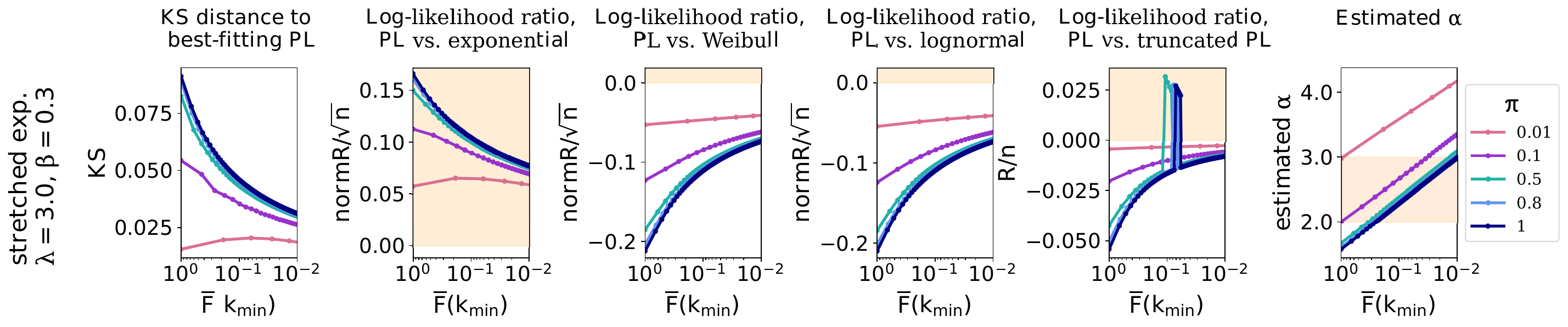}\\
\includegraphics[width=1\textwidth]{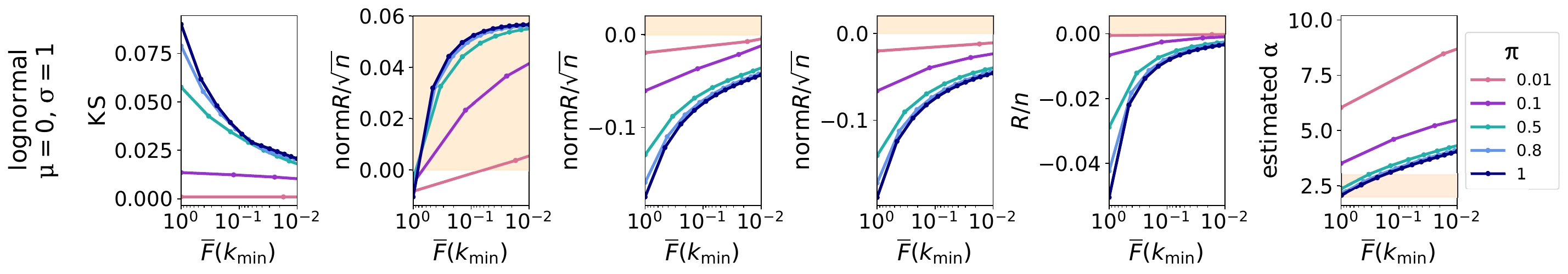}

\caption{Applying the estimators of the ML method to theoretical subsampled stretched exponential (upper row) and lognormal distribution (lower row) to assess how likely the power-law hypothesis can be expected to be rejected for different subsampling depths $\pi$ in the limit $n \to \infty$. 
The analytical limit values of the estimators are calculated by varying $k_{\mathrm{min}}$ but we present them as a function of the value of the CCDF at $k_{\mathrm{min}}$, $\bar{F}(k_{\mathrm{min}})$, to allow better comparison between distributions. Here, we define $\bar{F}(k_{\mathrm{min}}) = \sum_{k=k_\mathrm{min}} P(k)$. Note that the $x$-axis has been inverted to enhance readability; when moving to the right, we are considering larger and larger values of $k_\mathrm{min}$, which correspond to smaller values of $\bar{F}(k_{\mathrm{min}})$. The first column shows that the KS distance between the values of the distribution's CDF and the CDF of the best-fitting power-law distribution for a given $\bar{F}(k_{\mathrm{min}})$ tends to become smaller as the subsampling depth $\pi$ decreases, indicating that the goodness-of-fit test of the ML method can be expected to perform worse for smaller $\pi$. The next four columns show the results of the log-likelihood ratio tests where the power-law (PL) hypothesis is tested against an alternative hypothesis. The best-fitting parameters of the alternative distribution are obtained with the same optimization methods as in the implementation of Alstott \emph{et al.}\ \cite{alstott} using as initial parameter guesses the values of the Alstott \emph{et al.}\ implementation as $n \to \infty$. Depending on the alternative hypothesis, the results are expressed in terms of either the log-likelihood ratio $R$ or its normalized version $\mathrm{norm}R$ (details in SM IVA.3 \cite{supplemental_material}). If the ratio is positive (shaded yellow area), the incorrect PL model is favored over the alternative hypothesis whereas a negative ratio indicates that the alternative hypothesis is favored over the PL model. The log-likelihood ratio tests should be understood as a series; if all four tests favor the power-law hypothesis over the alternative distribution or give inconclusive results (ratio close to zero, not meaningful in the limit of large $n$), the power-law hypothesis is considered to get support. Consequently, for the considered subsampled lognormal and stretched exponential distributions, the power-law hypothesis is more likely to be correctly rejected the more negative the likelihood ratios are. The abrupt spikes in the fifth column result from suboptimal estimation of parameters of the truncated power-law distribution. The last column shows that for a given $\bar{F}(k_\mathrm{min})$, the estimated value of the power-law exponent $\alpha$ is expected to converge to a larger value for smaller subsampling depths. The shaded area marks the range to which the estimated $\alpha$ must fall for the criteria of strong evidence for the PL hypothesis to be met.}
\label{fig:analytical_clauset}
\end{figure*}

To get insight into the behaviour of the goodness-of-fit test, we calculate for each $k_{\mathrm{min}}$ the Kolmogorov-Smirnov distance between the subsampled distribution's CDF and the CDF of the best-fitting power law as $n \to \infty$.  
Our results show that the KS distance between the distribution's CDF and the CDF of the best-fitting power law for a given $\bar{F}(k_{\mathrm{min}})$ tends to become smaller as the subsampling depth decreases; however, this effect remains small for moderate subsampling depths and becomes less prominent when moving towards the tail of the distribution [Fig.~\ref{fig:analytical_clauset}, see also Fig.~9 in SM V \cite{supplemental_material} for results for distributions with other parameter combinations]. In general, a smaller KS distance indicates that the goodness-of-fit test is more likely to accept the power-law hypothesis. This is because the goodness-of-fit test proceeds by generating a number of bootstrapped samples from the fitted power-law model, after which a power-law model is fitted to each sample and the KS distance is calculated for each sample with respect to its own fitted model. The larger the fraction of the bootstrapped samples with KS distance exceeding that of the original sample, the more likely the power-law model is deemed to be. Consequently, assuming that the variances of the distributions of the KS statistics are approximately equal for different subsampling depths, our results indicate that if the values of the smallest considered degree $k_{\mathrm{min}}$ for two subsampling depths are chosen so that $\bar{F}(k_{\mathrm{min}})$ is approximately the same for both, the goodness-of-fit test is less likely to correctly reject the PL hypothesis for the lower subsampling depth. While a similar KS distance can often be obtained for the smaller subsampling depth by considering a larger part of the tail, this kind of compensation is not expected to happen in practice since the optimal $\hat{k}_{\mathrm{min}}$ for empirical or simulated data is chosen by minimizing the KS distance, which for all tested subsampling depths except for $\pi=0.01$ tends to decrease towards the tail of the distribution. 

The log-likelihood ratios comparing the likelihood of the power-law hypothesis to that of alternative distributions behave in a similar manner. With empirical data, the power-law hypothesis is considered to get support if none of the four alternative distributions is favored over the power-law model (i.e.\ the log-likelihood ratios are all positive or the corresponding $p$-values are non-significant). Hence, the more negative the theoretical log-likelihood ratios are, the more likely the power-law hypothesis will be correctly rejected. Our results on the theoretical distributions show that when the alternative hypothesis against the power-law model is either the lognormal, the Weibull, or the truncated power-law distribution, the limit value of the likelihood ratio tends to become less negative for smaller subsampling depths for a given $\bar{F}(k_{\mathrm{min}})$, meaning that the alternative hypothesis is less strongly favored over the power-law model. However, the effect for moderate subsampling depths is again small and the difference between the subsampling depths decreases as the considered fraction of the tail becomes smaller. The exponential distribution is the only alternative distribution with the opposite tendency: for the tested subsampled probability distributions, the power-law hypothesis tends to be favored over the exponential model for all subsampling depths, but the ratio tends closer to zero as the subsampling depth decreases. 

Finally, we observe that for a given $\bar{F}(k_{\mathrm{min}})$, the limit value of the estimated power-law exponent $\alpha$ is in general larger for smaller subsampling depths. For some tested subsampled distributions with $\pi= 0.01$, the distribution is not likely to fill the strong evidence criteria for the power-law hypothesis because the values of $\alpha$ lie above the allowed range for almost all values of $k_{\mathrm{min}}$. 

Overall, our results show that the goodness-of-fit test and the log-likelihood ratio tests lose some of their power in classifying the distribution’s type correctly as the subsampling depth decreases.

\section{\label{sec:level3}Discussion}

In this work, we have 
investigated how well the maximum likelihood method of \cite{clauset,broido}~and the extreme value method of \cite{voitalov}~succeed in recognizing power-law distributions when the data is heavily subsampled with the incident subsampling strategy. As subsampled power-law distributions have been shown to approach the original power law asymptotically, we hypothesized that the methods would continue to work on subsampled data as long as the sample sizes remained reasonable. With the strong evidence criteria of the ML method, however, suboptimal estimation of the beginning of the power-law tail led to a substantial false rejection rate of the power-law hypothesis for the subsamples. While the EV method correctly recognized subsampled power-law distributions, it sometimes misclassified subsamples from both lognormal and stretched exponential distributions  as power laws. However, these false positives tended to result not from the subsampling itself, but from the estimators' inability to classify the original sample correctly due to the underlying distribution converging too slowly to its asymptotic extreme value distribution. 

Interestingly, we observed that while especially the lognormal distribution started to visually resemble a power law as the subsampling depth decreased, subsampling seemed to enhance the performance of the EV method in correctly classifying lognormal and stretched exponential distributions. This effect was visible especially for the Hill estimator; the fraction of probability mass in the distribution's tail allowing for correctly classifying the distribution's type was in general larger for lower subsampling depths, in some cases to the extent that the expected absolute number of nodes in this part of the tail increased. The moments and the kernel estimators followed the same trend, but for many of the tested distributions, the estimators started to oscillate at very low subsampling depths ($\pi=0.01$), which resulted in the fraction getting seemingly smaller. 

Overall, our results imply that the classifications obtained with the EV method should be accepted with some caution if very heavy-tailed distributions (such as the lognormal with $\sigma = 1.3$ or the stretched exponential distribution with $\beta=0.3$) are valid alternatives for the power-law hypothesis. As noted already in \cite{malevergne2003}, a result that the data belongs to the MDA of the Gumbel distribution seems to be relatively reliable, while a classification to the MDA of the Fr\'echet distribution (thus supporting the power-law hypothesis) is not equally informative. It has been argued, however, that while it is often important to know whether a distribution is heavy-tailed, further identifying it as a power-law distribution may not bring any considerable additional value \cite{critical_truths}. In some cases distinguishing between a power-law and a lognormal distribution might simply not be important, or alternatively the lognormal and the stretched exponential distributions might not be relevant candidates for the question at hand.  Consequently, if the main interest lies instead in e.g.\ determining whether a sample originates from an exponential distribution or from a power-law distribution, the EV method may be a suitable alternative even if the data is heavily subsampled. 

While the EV method is at times too permissible, the strong evidence criteria of the ML method avoid this drawback with the cost of an increased false rejection rate. Especially the requirement of the exponent $\alpha$ staying in the range $[2,3]$ results in many false rejections for power laws with $\alpha$ close to the limits of this range. The simulations showed that most subsamples from a pure power law did not exhibit strong evidence for the power-law hypothesis due to suboptimal estimation of the start of the power-law tail. In addition, we showed that distinguishing the alternative distributions from power laws using the goodness-of-fit and the log-likelihood ratio tests of the ML method becomes increasingly difficult for lower subsampling depths. It is important to remember, however, that these results apply directly only to incident subgraph sampling, and other subsampling methods might produce substantially different results.

In general, while the automatic determination of the fraction of the tail considered in the analysis has it benefits -- including the fact that no subjective determination of the threshold is needed -- one should not trust this estimate blindly (a point raised with regard to the ML method already by e.g.\ Refs.~\cite{voitalov} and \cite{corral}). At the very least, it might be useful to examine more closely the range of values of $k_{\mathrm{min}}$ for which certain conclusions are valid; examining how the estimates change as a function  of $k_{\mathrm{min}}$ might in some cases even offer further insight into the distribution's type as shown in Ref.~\cite{salje}. The automatic estimation is especially likely to fail if the probability mass of the distribution is heavily concentrated on the small degrees, from which the PDF decays in a seemingly convex manner on a log-log-plot. However, as noted by Stumpf \emph{et al.}~\cite{stumpf_scalefree}, this kind of convex decay on a log-log plot is not commonly observed in real-world networks, and the problem might thus be overly pronounced in the simulated subsamples.


Overall, while our results highlight the importance of analyzing the same issue with different approaches, even using the two methods in combination does not always allow one to deduce with reasonable confidence whether a subsample originates from a power-law distribution. Naturally, the situation is likely to be even more complicated when analyzing real-world networks with more noise and variation. Consequently, assessing whether other methods -- such as the maximum entropy test of Bee \emph{et al.}~\cite{bee2011}, the Wilk's test used in Ref.~\cite{malevergne2003}, the finite size scaling method of Ref.~\cite{serafino} or the approaches presented in Refs.~\cite{zhang}, \cite{corral2018} and \cite{artico} --  could fruitfully complement the methods addressed in this work remains a task for future research.

\section*{Acknowledgments}

We acknowledge the computational resources provided by the Aalto Science-IT project.

\InputIfFileExists{bibliography.bbl}

\clearpage



\section*{Supplementary material}

# Supplemental material

Silja Sormunen,[1] Lasse Leskelä,[2] and Jari Saramäki[1]
[1]*Department of Computer Science, Aalto University, 00076 Espoo, Finland*
[2]*Department of Mathematics and Systems Analysis, Aalto University, 00076 Espoo, Finland*

## I. MODIFICATIONS TO THE IMPLEMENTATIONS

### A. Maximum likelihood method

As explained in Section IIIA of the article, we use a combination of the implementations by Alstott *et al.* [1] and Broido and Clauset [2] when applying the ML method to the simulated subsamples. For the goodness-of-fit test, we use the *plpval* function by Broido and Clauset where we have modified the code so that the model-fitting is done with the *powerlaw* package by Alstott *et al.* To be precise, we have replaced the line

```
[newalpha, newxmin, newntail, newLpl, newgof] = pl(newx)
```

with

```
results_bootstrapped = powerlaw.Fit(newx, discrete=True, estimate_discrete = False)
newalpha, newxmin, newgof = float(results_bootstrapped.power_law.alpha),  \
        int(results_bootstrapped.power_law.xmin), results_bootstrapped.power_law.KS(newx)
```

In addition, we have updated the code from Python2 to Python3 by simply adding parentheses to the *print* functions.

We have chosen to use a combination of the two implementations due to an issue we encountered in the implementation by Broido and Clauset regarding the calculation of Hurwitz zeta function. When choosing the optimal value of $k_{\min}$, a power-law model is fitted using each of the unique values of $k$ as $k_{\min}$ as shown below:

```
    # loop over the xmin values at find the best fit at each
    for i in range(len(xminV)):
        xmin = xminV[i]
        xtail = x[x>=xmin]
        ntail = len(xtail)
        # optimize over alpha
        # find the corresponding array of conditional log likelihoods
        Ls = -alphaV*np.sum(np.log(xtail)) - ntail*np.log(constV)
        # pull out the location of the best fit alpha
        aind = Ls.argmax()
        # find what alpha value is at this index
        alpha = alphaV[aind]
        # compute the KS statistic
        # theoretical cdf
        cdf = np.cumsum(range(np.min(xtail), np.max(xtail)+1)**(-alpha)/constV[aind])
        #  binned data
        xhist = np.histogram(xtail,range(np.min(xtail), np.max(xtail)+2))
        # empirical cdf
        edf = np.cumsum(xhist[0])/float(ntail)
        # KS stat
        ks = np.max(np.abs(cdf-edf))
        # add this KS stat and alpha to the array of fits
        fitV[i] = np.array([ks, alpha])
        # update the constants
        for j in range(xmin-xminprev):
            constV += -(xmin+j)**(-alphaV)
        xminprev = xmin
```

This approach works if all integers between the minimum and maximum value in the sample are present in the data, in other words if for each unique value of $k_i$ present in the data it is true that $k_i = k_{i-1} + 1$. However, if this is not the case, the updating of the Hurwitz zeta function (given by the variable **constV**) will not work correctly, resulting in a possibly inaccurate estimation of the optimal $\hat{k}_{\min}$. This issue can be fixed by removing the lines

```
for j in range(xmin-xminprev):
    constV += -(xmin+j)**(-alphaV)
```

and updating **constV** earlier in the loop so that the beginning of the loop becomes *e.g.*

```
for i in range(len(xminV)):
    xmin = xminV[i]
    xtail = x[x>=xmin]
    ntail = len(xtail)

    if not xminprev == min(xminV) - 1:
        for j in range(xmin-xminprev):
            constV += -(xminprev+j)**(-alphaV)
```

### B. Extreme value method

As explained in Section IIB of the article, the code of Voitalov *et al.* [3] employs $\lfloor nh \rfloor - 1$ as the upper limit in the summation formula of the kernel estimator (see Eqs. 9, 11 and 13 of the article), while we use $\lfloor nh \rfloor$ as the upper limit (except when $h = 1$, as explained below). We illustrate the reason for this change using the first term of the kernel estimator,

$$\hat{\xi}_{h,n}^{\text{pos}} = -\int_0^h u K_h(u) \, d \ln Q_n(1-u) \tag{1}$$

$$= \sum_{i=1}^{\lfloor nh \rfloor} \frac{i}{n} K_h\left(\frac{i}{n}\right) \ln\left(\frac{X_{(i)}}{X_{(i+1)}}\right), \tag{2}$$

where $K_h(u) = \frac{15}{8h}(1 - (\frac{u}{h})^2)^2$, $Q_n$ denotes the empirical quantile function, $X_{(i)}$ is the $i$th largest data point in the sample, and $h$ is fraction of the tail to be considered in the calculation.

Let us first assume that $h$ is not a multiple of $\frac{1}{n}$, in which case Eq. 1 can be expressed as

$$\hat{\xi}_{h,n}^{\text{pos}} = \sum_{i=1}^{\lfloor nh \rfloor} \frac{i}{n} K_h\left(\frac{i}{n}\right) \left[\ln Q_n\left(1 - \frac{i-1}{n}\right) - \ln Q_n\left(1 - \frac{i}{n}\right)\right] + h K_h(h) \left[\ln Q_n\left(1 - \frac{\lfloor nh \rfloor}{n}\right) - \ln Q_n(1-h)\right]. \tag{3}$$

Since $K_h(h) = 0$, the last term in the above equation equals zero, and we are left with the sum from 1 up to $\lfloor nh \rfloor$. We note, however, that if $h$ is a multiple of $\frac{1}{n}$, we have that $\lfloor nh \rfloor = nh$, and since the last term in the sum corresponding to $i = \lfloor nh \rfloor$ again equals zero due to $K_h(h) = 0$, summing up to $\lfloor nh \rfloor$ is equal to summing up to $\lfloor nh \rfloor - 1$. Hence, if $h$ happens to be a multiple of $\frac{1}{n}$, these two upper limits yield identical results. The two upper limits give the same results also if $Q_n(1 - \frac{\lfloor nh \rfloor - 1}{n}) = Q_n(1 - \frac{\lfloor nh \rfloor}{n})$. However, if neither of these two conditions are met, the values obtained using $\lfloor nh \rfloor - 1$ as the upper limit are inaccurate (see Fig. 1 for illustration).

To correct the upper limit in the code of Ref. [3], we have modified the functions **get_biweight_kernel_estimates** and **get_triweight_kernel_estimates** by replacing the line

```
max_i_vector = (np.floor((n*h_arr))-2.).astype(int)
```

with

```
max_i_vector = (np.floor((n*h_arr))-1.).astype(int)
max_i_vector[-1] = max_i_vector[-1]-1
```

The last line is necessary since for $h = 1$ we need to use the upper limit $\lfloor nh \rfloor - 1 = n - 1$ as there are only $n$ data points. As reasoned above, using $\lfloor nh \rfloor - 1$ as the upper limit gives correct results in this case since $h = 1$ is a multiple of $1/n$.

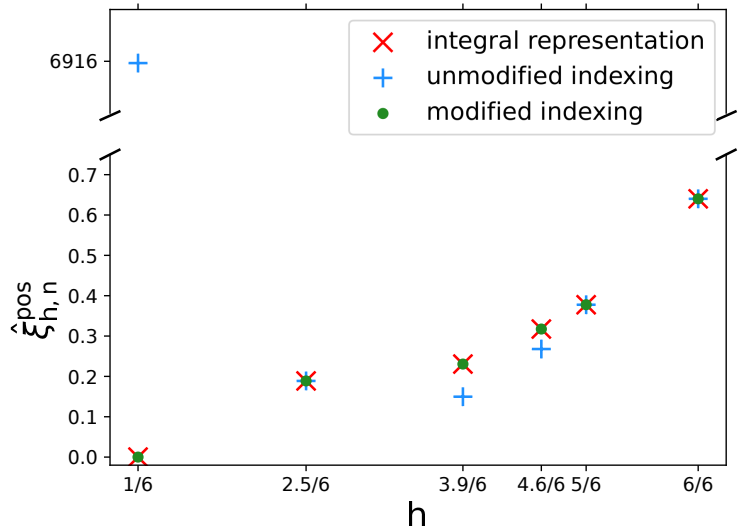

FIG. 1. Values of $\hat{\xi}_{h,n}^{\text{pos}}$ as a function of the fraction $h$ of the considered tail for a small arbitrarily chosen data set $[10, 7, 7, 5, 3, 1]$. The red crosses display the values of $\hat{\xi}_{h,n}^{\text{pos}}$ obtained by numerically integrating the expression in Eq. 28 using the **quad** function of the **scipy** library. The blue markers show the values obtained with the unmodified code (using $\lfloor nh \rfloor - 1$ as the upper limit) and the green markers the results after the indexing has been corrected (using $\lfloor nh \rfloor$ as the upper limit). We notice that the original indexing yields correct results when $h$ is a multiple of $1/n$ except when $h = 1/n$, for which the value diverges heavily since the code gives $-1$ as the corresponding index, meaning that $n-1$ is accidentally used as the upper limit in the summation. The original indexing yields correct results also for $h = 2.5/6$, since $Q_n(1 - \frac{\lfloor nh \rfloor - 1}{n}) = 7$ and $Q_n(1 - \frac{\lfloor nh \rfloor}{n}) = 7$ are equal. For other values of $h$, the values produced by the original code are not accurate.

## II. ESTIMATION ACCURACY OF THE POWER-LAW EXPONENT

Fig. 2 displays how the estimated power-law exponent $\hat{\alpha}$ and the threshold $\hat{k}_{\min}$ change as the subsampling depth $\pi$ decreases for power-law distributions with varying exponents $\alpha$. The median size of the subsamples as well as the median maximum degree in the subsamples are shown in Fig. 4. In Figs. 2 and 4 the size of the network before subsampling is $n_0 = 10^5$. Figs. 3 and 5 show the same results for networks with $n_0 = 10^6$.

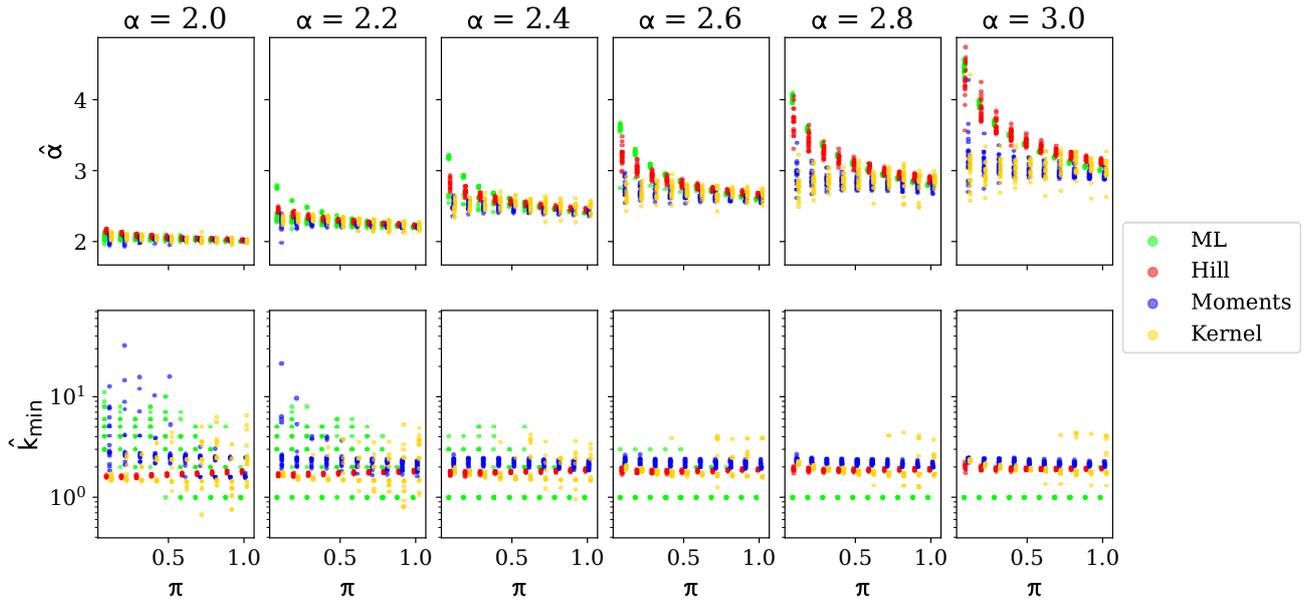

FIG. 2. Estimates of the power-law exponent $\hat{\alpha}$ and the threshold $\hat{k}_{\min}$ for different subsampling depths $\pi$. Each color corresponds to one estimator. The title of each column shows the true power-law exponent $\alpha$ of the original distribution. The number of data points before subsampling is $10^5$.

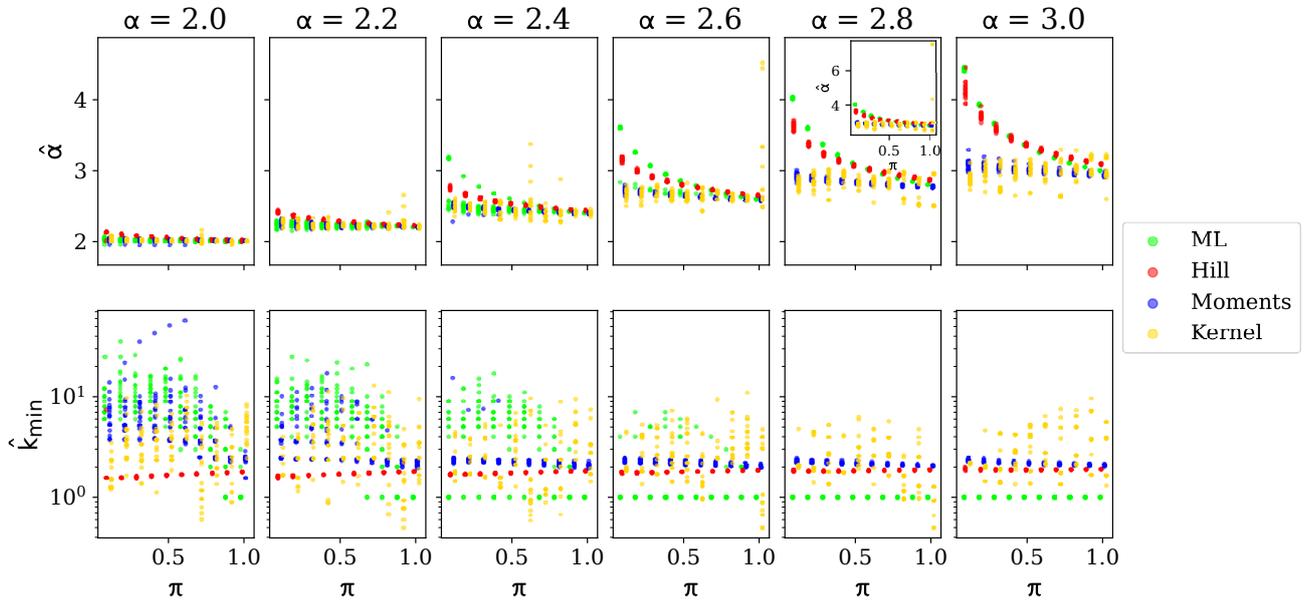

FIG. 3. Estimates of the power-law exponent $\hat{\alpha}$ and the threshold $\hat{k}_{\min}$ for different subsampling depths $\pi$. The number of data points before subsampling is $10^6$. The axis limits are the same as in Fig. 2. There are two outlier values of $\hat{\alpha}$ for $\alpha = 2.8$; the whole data with the uncut axis is shown in the inset.

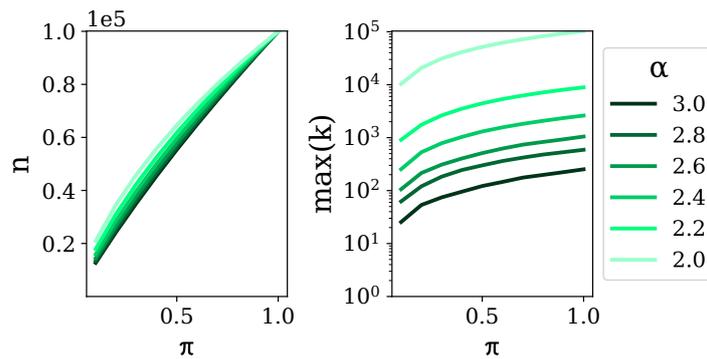

FIG. 4. The median size $n$ and the median maximum degree in subsamples from distributions with varying power-law exponents $\alpha$ and $n_0 = 10^5$

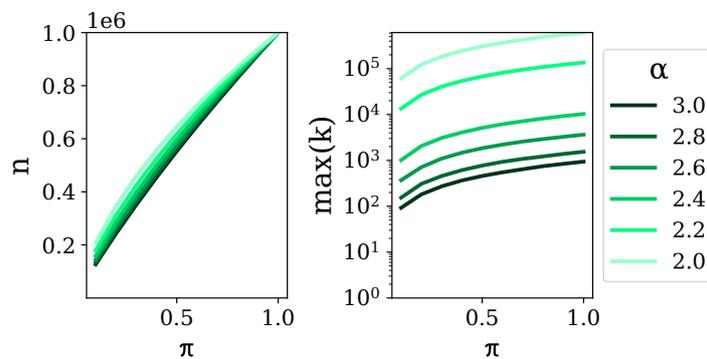

FIG. 5. The median size $n$ and the median maximum degree in subsamples from distributions with varying power-law exponents $\alpha$ and $n_0 = 10^6$

### III. CLASSIFICATION ACCURACY OF INDIVIDUAL ESTIMATORS

Fig. 6 displays the fraction of (sub)samples classified as power laws by the three EV estimators (see Figs. 3 and 4 of the article, which show the same results for the three estimators combined).

The performance of individual criteria of the ML method (see the list of criteria in Section IIA of the article) and some of their combinations are shown in Fig. 7. We note that the criterion about the number of nodes in the considered tail being at least 50 (criterion 2) is fulfilled by almost all (sub)samples and hence not displayed in the the figure. Interestingly, we observe that the log-likelihood criterion (criterion 4) works in general better than the goodness-of-fit test (criterion 3); the fraction of correctly classified subsamples from power-law distributions is higher while the rate of false positives is in general lower. Comparing the log-likelihood ratio criteria with the *some evidence* criteria (requires either criteria 3 or 4 fulfilled, see Figs. 3 and 4 of the article), we observe that the performance for subsamples from power laws and the BA model is slightly better for the *some evidence* criteria, but this difference is small. However, performance on alternative distributions is clearly better when using only the log-likelihood ratio criteria.

Furthermore, we notice that leaving the goodness-of-fit criterion out of the *strong evidence* criteria (criteria 1+2+3+4, see Figs. 3 and 4 of the article) results in better performance for subsamples from power-law distributions as well as the BA model, but the number of false positives for the lognormal distribution with $\sigma = 2$ is slightly higher. All other alternative distributions are correctly classified as non-power-laws by both combinations.

Finally, we observe that omitting the criterion about the estimated power-law exponent falling between 2 and 3 (criterion 1) from the *strong evidence* criteria does not substantially increase the rate of correctly classified power laws, while an increased amount of subsamples from alternative distributions are incorrectly classified as power laws. Hence, excluding this criterion is not advisable.

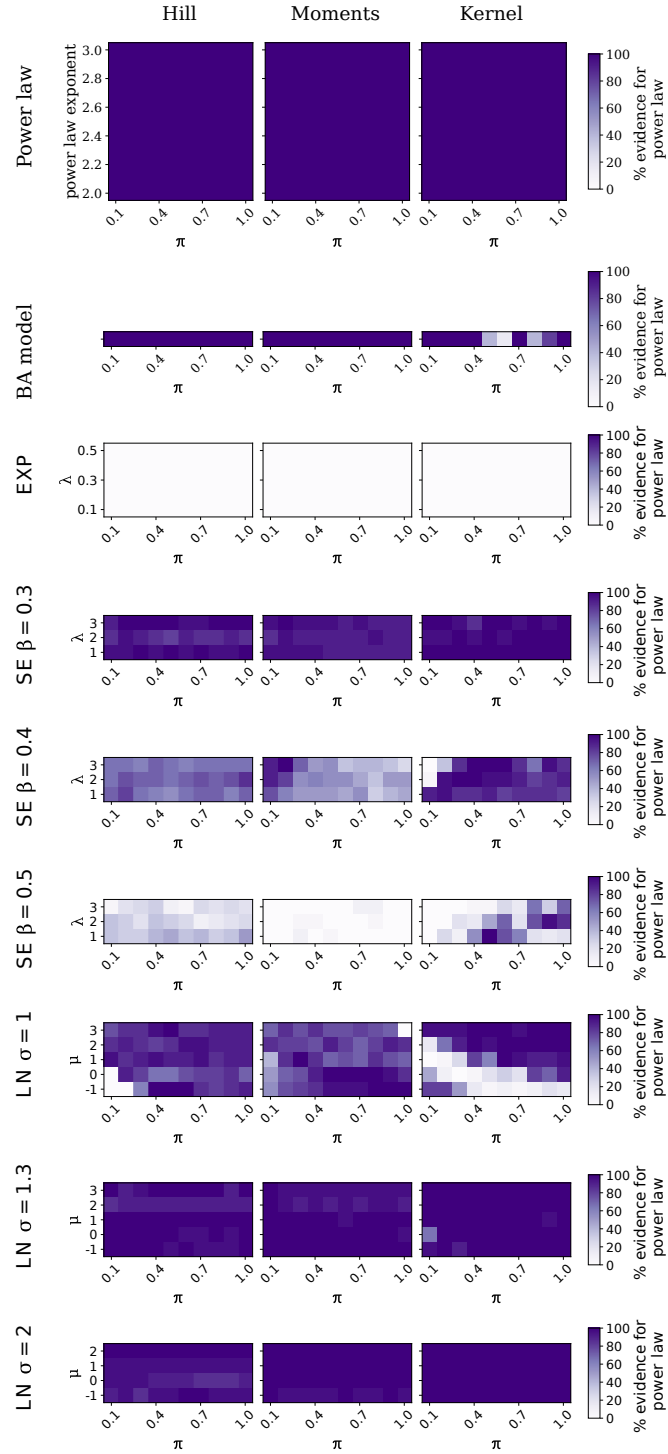

FIG. 6. Performance of the two methods with samples from power law (PL) and the Barabási-Albert (BA) model as well as from the exponential (EXP), stretched exponential (SE) and the lognormal (LN) distributions. The darker the color, the higher the fraction of subsamples classified as power laws (for the power law and the BA model, this indicates a correct classification while for the other distributions this would be a misclassification).


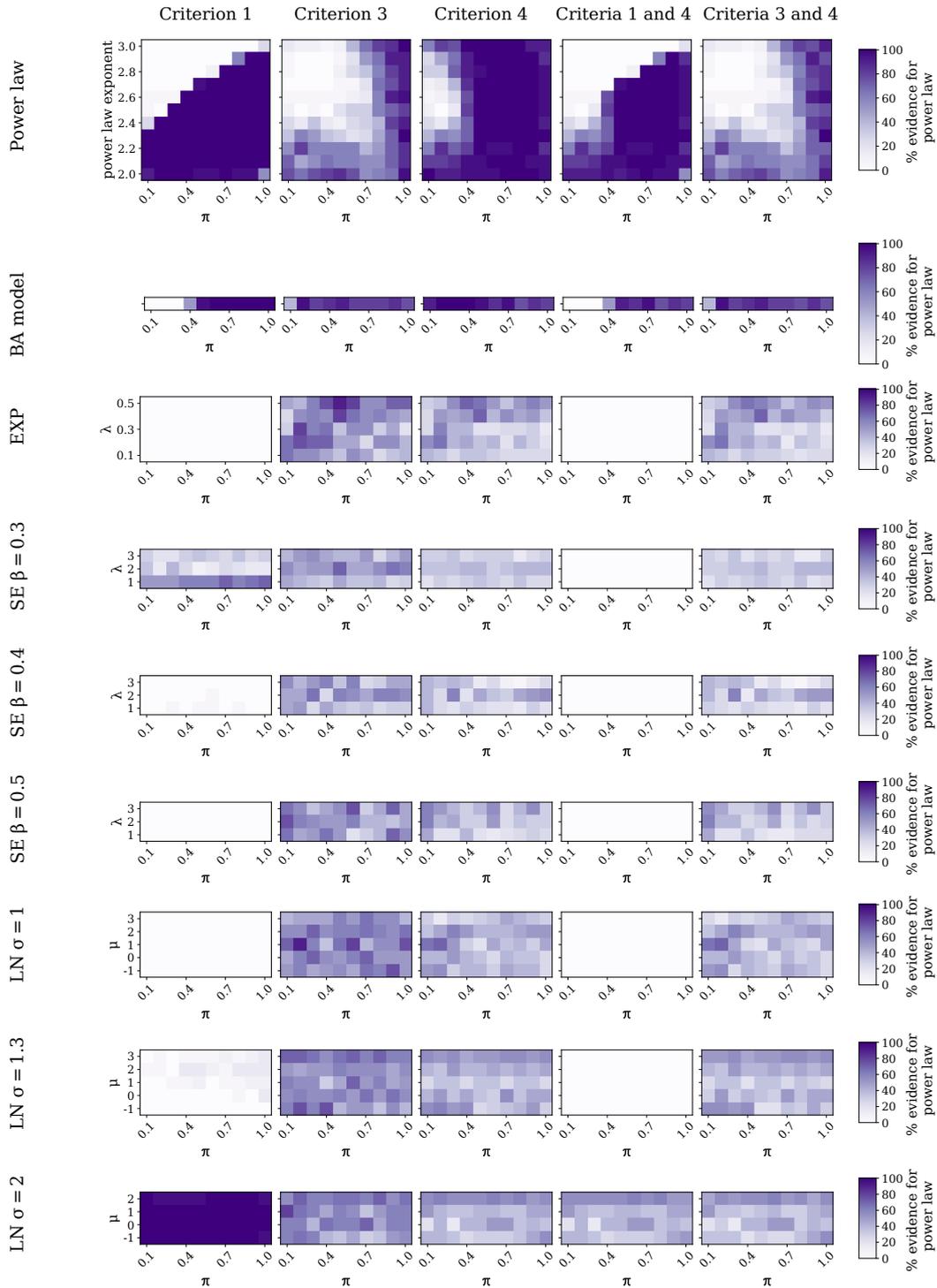

FIG. 7. Performance of different combinations of tests included in the ML method. The numbering of criteria corresponds to that in Section IIA of the article: 1. estimated power-law exponent is between 2 and 3; 2. $n_{\text{tail}} \geq 50$; 3. goodness-of-fit test does not reject the PL hypothesis; 4. no alternative distribution is preferred in the log-likelihood ratio tests over PL. The second criteria is fulfilled by almost all subsamples and hence left out of the figures. The combination of criteria 1 and 3 is likewise omitted as it produces worse results than the combinations displayed here. The darker the color, the higher the fraction of subsamples classified as power laws. The percentages are calculated over 20 samples with the number of data points before subsampling equal to $10^5$.

## IV. LIMIT VALUES OF THE ESTIMATORS

### A. Maximum likelihood method

In this section we derive the limiting values of the estimators of the maximum likelihood (ML) method [2, 4] as $n \to \infty$ for a given $k_{\min}$ and an arbitrary probability distribution with a discrete probability density function $P(k)$ and cumulative distribution function $F(k)$. As in the main text, $k_{\min}$ denotes the smallest degree $k$ included in the estimation. We denote the normalized probabilities $P(k)/\sum_{k \geq k_{\min}} P(k)$ by $P^{\text{tail}}(k)$ and the corresponding values of the CDF by $F^{\text{tail}}(k)$. As the value of $k_{\min}$ is fixed, the number of data points $k_i \geq k_{\min}$ approaches infinity as $n \to \infty$. For convenience, we denote the number of data points $k_i \geq k_{\min}$ by $n$ in the following.

#### 1. ML estimate of power-law exponent

The maximum likelihood estimate of the power-law exponent is obtained by maximizing the logarithm of the likelihood function with respect to $\alpha \in (1, \infty)$ [4],

$$\hat{\alpha}_n = \underset{\alpha}{\operatorname{argmax}}\, L(\alpha) = \underset{\alpha}{\operatorname{argmax}} \left\{ -\ln\left[\zeta(\alpha, k_{\min})\right] n - \alpha \sum_{i=1}^{n} \ln(k_i) \right\}, \tag{4}$$

where the summation runs over data points in the considered tail, i.e. such that $k_i \geq k_{\min}$. By the strong law of large numbers, for a given $k_{\min}$ and $\alpha$,

$$\frac{1}{n} L(\alpha) \xrightarrow{a.s.} -\ln[\zeta(\alpha, k_{\min})] - \alpha\, \mathbb{E}[\ln(k) \mid k \geq k_{\min}], \tag{5}$$

where *a.s.* stands for convergence almost surely. Consequently, since the convergence is uniform over the values of $\alpha$,

$$\hat{\alpha}_n \xrightarrow{a.s.} \hat{\alpha} = \underset{\alpha}{\operatorname{argmax}} \left\{ -\ln[\zeta(\alpha, k_{\min})] - \alpha\, \mathbb{E}[\ln(k) \mid k \geq k_{\min}] \right\}. \tag{6}$$

We can confirm that the above function is indeed maximized at the unique parameter value $\hat{\alpha}$ by analyzing the Kullback-Leibler (KL) divergence

$$D_{\text{KL}}(P^{\text{tail}} \| P^{\text{tail}}_{\text{PL},\alpha}) = \sum_{k \geq k_{\min}} P^{\text{tail}}(k) \ln P^{\text{tail}}(k) + \ln \zeta(\alpha, k_{\min}) + \alpha \sum_{k \geq k_{\min}} P^{\text{tail}}(k) \ln k, \tag{7}$$

where $P^{\text{tail}}_{\text{PL},\alpha} = \zeta(\alpha, k_{\min})^{-1} k^{-\alpha}$. In the limit of large $n$, maximizing the logarithm of the likelihood function is equivalent to minimizing the KL divergence. Differentiating the Hurwitz zeta function term by term shows that

$$\frac{\partial}{\partial \alpha} D_{\text{KL}}(P^{\text{tail}} \| P^{\text{tail}}_{\text{PL},\alpha}) = \sum_{k \geq k_{\min}} P^{\text{tail}}(k) \ln k - \zeta(\alpha, k_{\min})^{-1} \sum_{k \geq k_{\min}} k^{-\alpha} \ln k \tag{8}$$

$$= \sum_{k \geq k_{\min}} P^{\text{tail}}(k) \ln k - \sum_{k \geq k_{\min}} P^{\text{tail}}_{\text{PL},\alpha}(k) \ln k, \tag{9}$$

and

$$\frac{\partial^2}{\partial \alpha^2} D_{\text{KL}}(P^{\text{tail}} \| P^{\text{tail}}_{\text{PL},\alpha}) = \frac{\partial^2}{\partial \alpha^2} \ln \zeta(\alpha, k_{\min}) = \frac{\zeta''(\alpha, k_{\min})}{\zeta(\alpha, k_{\min})} - \left(\frac{\zeta'(\alpha, k_{\min})}{\zeta(\alpha, k_{\min})}\right)^2 \tag{10}$$

$$= \sum_{k \geq k_{\min}} P^{\text{tail}}_{\text{PL},\alpha}(k) \ln^2 k - \left(\sum_{k \geq k_{\min}} P^{\text{tail}}_{\text{PL},\alpha}(k) \ln k\right)^2. \tag{11}$$

We see that $\frac{\partial^2}{\partial \alpha^2} D_{\text{KL}}(P^{\text{tail}} \| P^{\text{tail}}_{\text{PL},\alpha})$ equals the variance of $\ln X$ with $X$ sampled from $P^{\text{tail}}_{\text{PL},\alpha}$, and therefore $\alpha \mapsto D_{\text{KL}}(P^{\text{tail}} \| P^{\text{tail}}_{\text{PL},\alpha})$ is strictly convex on $(1, \infty)$. Furthermore, we note that $\lim_{\alpha \downarrow 1} = \infty$, so that $\lim_{\alpha \downarrow 1} D_{\text{KL}}(P^{\text{tail}} \| P^{\text{tail}}_{\text{PL},\alpha}) = \infty$. We also note that $\lim_{\alpha \uparrow \infty} \frac{\partial}{\partial \alpha} D_{\text{KL}}(P^{\text{tail}} \| P^{\text{tail}}_{\text{PL},\alpha}) = \sum_{k \geq k_{\min}} P^{\text{tail}}(k) \ln k - \ln k_{\min}$. We conclude that whenever $f$ is nondegenerate (i.e. $P^{\text{tail}}(k_{\min}) < 1$), the function $\alpha \mapsto D_{\text{KL}}(P^{\text{tail}} \| P^{\text{tail}}_{\text{PL},\alpha})$ on $(1, \infty)$ is minimized at the unique parameter value at which $\frac{\partial}{\partial \alpha} D_{\text{KL}}(P^{\text{tail}} \| P^{\text{tail}}_{\text{PL},\alpha}) = 0$, that is,

$$\sum_{k \geq k_{\min}} P^{\text{tail}}_{\text{PL},\alpha}(k) \ln k = \sum_{k \geq k_{\min}} P^{\text{tail}}(k) \ln k. \tag{12}$$

### 2. Kolmogorov-Smirnov distance

The Kolmogorov-Smirnov distance between the empirical distribution function $F_n^{\text{tail}}(k)$ and the fitted power-law model $F_{PL,\hat{\alpha}_n}^{\text{tail}}(k)$ on the degrees $k \geq k_{\min}$ is given by [4]

$$d_{\text{KS}}(F_n^{\text{tail}}, F_{\text{PL},\hat{\alpha}_n}^{\text{tail}}) = \max_{k \geq k_{\min}} |F_n^{\text{tail}}(k) - F_{\text{PL},\hat{\alpha}_n}^{\text{tail}}(k)|. \tag{13}$$

Denoting by $F_{\text{PL},\hat{\alpha}}^{\text{tail}}$ the distribution function of a power-law distribution with $\hat{\alpha}$ obtained using Eq. 6, we obtain the triangle inequality

$$\left| d_{\text{KS}}(F_n^{\text{tail}}, F_{\text{PL},\hat{\alpha}_n}^{\text{tail}}) - d_{\text{KS}}(F^{\text{tail}}, F_{\text{PL},\hat{\alpha}}^{\text{tail}}) \right| \leq d_{\text{KS}}(F_n^{\text{tail}}, F^{\text{tail}}) + d_{\text{KS}}(F_{\text{PL},\hat{\alpha}_n}^{\text{tail}}, F_{\text{PL},\hat{\alpha}}^{\text{tail}}). \tag{14}$$

The Glivenko-Cantelli theorem ensures that $d_{\text{KS}}(F_n^{\text{tail}}, F^{\text{tail}}) \xrightarrow{a.s.} 0$. Similarly, $d_{\text{KS}}(F_{\text{PL},\hat{\alpha}_n}^{\text{tail}}, F_{\text{PL},\hat{\alpha}}^{\text{tail}}) \xrightarrow{a.s.} 0$ since $\hat{\alpha}_n \xrightarrow{a.s.} \hat{\alpha}$ (Eq. 6) and $\hat{\alpha} \mapsto F_{\text{PL},\hat{\alpha}}^{\text{tail}}$ is continuous in the KS metric. Hence, we can conclude that

$$d_{\text{KS}}(F_n^{\text{tail}}, F_{\text{PL},\hat{\alpha}_n}^{\text{tail}}) \xrightarrow{a.s.} d_{\text{KS}}(F^{\text{tail}}, F_{\text{PL},\hat{\alpha}}^{\text{tail}}). \tag{15}$$

### 3. Log-likelihood ratio

The log-likelihood ratio comparing the likelihood of the power-law hypothesis to that of an alternative hypothesis is given by

$$R = \sum_{i=1}^n \left[ \ln(P_{\text{PL},\hat{\alpha}_n}^{\text{tail}}(k_i)) - \ln(P_{\text{Alt},\hat{\theta}_n}^{\text{tail}}(k_i)) \right], \tag{16}$$

where $P_{\text{PL},\hat{\alpha}_n}^{\text{tail}}(k)$ gives the PDF of the fitted power-law distribution and $P_{\text{Alt},\hat{\theta}_n}^{\text{tail}}(k)$ that of an alternative distribution with ML-estimated parameter(s) $\hat{\theta}_n$. To conclude whether the log-likelihood ratio differs significantly from zero, the ML method employs a normalized log-likelihood ratio given by

$$\text{norm}R = \frac{R}{\hat{\sigma}_n \sqrt{n}}, \tag{17}$$

where $\hat{\sigma}_n$ denotes the standard deviation of the individual summation terms in Eq. 16.

Vuong et al. [5] have shown that $\frac{1}{n}R \xrightarrow{a.s.} \mathbb{E}[\ln(P_{\text{PL},\hat{\alpha}}^{\text{tail}}(k)) - \ln(P_{\text{Alt},\hat{\theta}}^{\text{tail}}(k))]$ when the parameters of the fitted models are ML estimates and the fitted model distributions satisfy some general regularity conditions (Lemma 3.1 in [5]). In addition, $\hat{\sigma}_n \xrightarrow{a.s.} \sigma$ (Lemma 4.4 in [5]). Consequently, we obtain:

$$\frac{\text{norm}R}{\sqrt{n}} = \frac{1}{n\hat{\sigma}_n}R \xrightarrow{a.s.} \frac{1}{\sigma}\mathbb{E}[\ln(P_{\text{PL},\hat{\alpha}}^{\text{tail}}(k)) - \ln(P_{\text{Alt},\hat{\theta}}^{\text{tail}}(k))]. \tag{18}$$

The above expression applies only when comparing two non-nested distributions. If one of the distributions belongs to a subclass of the other distribution family, the fraction $R/\sigma_n$ tends to $0/0$, which is why the result of the comparison needs to be determined directly from the value of $R$ and the associated p-value (see [4]). Consequently, we examine the behaviour of $R/n$ when comparing the power-law hypothesis to the truncated power-law hypothesis.

As both $\text{norm}R/\sqrt{n}$ and $R/n$ depend on the sample size $n$, we cannot directly assess whether the value of $\text{norm}R$ should be considered significant or not. Still, examining the behaviour of this expression is useful in assessing the effect of subsampling depth on the performance of the log-likelihood ratio test.

### B. Extreme value method

In the following, we derive the limiting values of the estimators of the extreme value (EV) method [3] as $n \to \infty$ for a given $k_{\min}$. As explained in the main text, to mirror the empirical analysis as closely as possible, we assume the PDFs to be step functions where the probability $P(k)$ stays constant from $k - 0.5$ up to $k + 0.5$ for each integer $k \geq 1$.

### 1. Hill estimator

For a fixed $k_{\min}$, the Hill estimator gives an empirical estimate of the average log-excess over $k_{\min}$, in other words the sample mean of $[\ln(k_i) - \ln(k_{\min}) \mid k_i > k_{\min}]$. As $n \to \infty$, the Hill estimator for a given $k_{\min}$ converges by the strong law of large numbers almost surely to its expected value

$$\mathbb{E}[\hat{\xi}_{k_{\min}}^{\text{Hill}}] = \frac{\int_{k_{\min}}^{\infty} P(k)[\ln(k) - \ln(k_{\min})] \, dk}{\bar{F}(k_{\min})}. \tag{19}$$

We replace the upper limit of the integral with the large but finite limit used in the generation of the theoretical probability distributions. In practice, we obtain the integral of the numerator by dividing the support to blocks of length one, solving analytically the definite integral, and summing the results to give the expected value.

### 2. Moments estimator

The moments estimator is given by

$$\hat{\xi}_{\kappa,n}^{\text{Moments}} = \hat{\xi}_{\kappa,n}^{\text{Hill}} + 1 - \tfrac{1}{2}\left(1 - \frac{(\hat{\xi}_{\kappa,n}^{\text{Hill}})^2}{\hat{\xi}_{\kappa,n}^{\text{Hill},2}}\right)^{-1}.$$

In the limit $n \to \infty$, the moments estimator for a given $k_{\min}$ converges a.s. to

$$\mathbb{E}\big[\hat{\xi}_{k_{\min}}^{\text{Hill}}\big] + 1 - \frac{1}{2} \frac{\mathbb{E}\big[\hat{\xi}_{k_{\min}}^{\text{Hill},2}\big]}{\mathbb{E}\big[\hat{\xi}_{k_{\min}}^{\text{Hill},2}\big] - \mathbb{E}\big[\hat{\xi}_{k_{\min}}^{\text{Hill}}\big]^2}, \tag{20}$$

where

$$\mathbb{E}[\hat{\xi}_{k_{\min}}^{\text{Hill},2}] = \frac{\int_{k_{\min}}^{\infty} P(k)[\ln(k) - \ln(k_{\min})]^2 dk}{\bar{F}(k_{\min})}. \tag{21}$$

### 3. Kernel estimator

The kernel estimator is given by

$$\hat{\xi}_{h,n}^{\text{Kernel}} = \hat{\xi}_{h,n}^{\text{pos}} - 1 + \frac{\hat{q}_{h,n}^{(2)}}{\hat{q}_{h,n}^{(1)}}, \text{ where} \tag{22}$$

$$\hat{\xi}_{h,n}^{\text{pos}} = -\int_0^h u K_h(u) \, d \ln Q_n(1-u), \tag{23}$$

$$\hat{q}_{h,n}^{(1)} = -\int_0^h u^\gamma K_h(u) \, d \ln Q_n(1-u), \tag{24}$$

$$\hat{q}_{h,n}^{(2)} = -\int_0^h \frac{d}{du}[u^{\gamma+1} K_h(u)] \, d \ln Q_n(1-u), \tag{25}$$

where $h$ is a bandwidth parameter determining the fraction of considered nodes, $K_h(u) = \frac{15}{8h}(1-(\frac{u}{h})^2)^2$, and the parameter $\beta > 0.5$ [6]. The limit as $n \to \infty$ for a given bandwidth $h$ is consequently obtained by replacing the empirical quantile function $Q_n$ of the random variable $X$ with the true quantile function $Q$. We use the term $\hat{\xi}_{h,n}^{\text{pos}}$ to illustrate that this interchange of the limit and integral operator is allowed assuming that $\mathbb{E}[X]$ is finite.

First, we convert the Lebesgue-Stieltjes integral to a Lebesgue integral using integration by parts using the observation that $K_h(h) = 0$,

$$\hat{\xi}_{h,n}^{\text{pos}} = -\int_0^h u K_h(u) d \ln Q_n(1-u) \tag{26}$$

$$= -h K_h(h) \ln Q_n(1-h) + 0 + \int_0^h \ln Q_n(1-u) d[u K_h(u)] \tag{27}$$

$$= \int_0^h \ln Q_n(1-u) \frac{d}{du}[u K_h(u)] \, du, \tag{28}$$

where the last step is allowed as the derivative of $uK_h(u)$ is differentiable on $(0,h)$ with derivative integrable on $[0,h]$. The interchange of limit and integral is allowed for a Lebesgue integral if the integrand converges in $\mathcal{L}^1$. It is known that if $\mathbb{E}[X]$ is finite, then $Q_n \to Q$ in $\mathcal{L}^1$ almost surely [7]. On the other hand, we know that $\mathcal{L}^1$-convergence is preserved in compositions with uniformly continuous functions as well as with bounded functions (Theorem 5, Corollary 2 in Ref. [8]). Consequently, since the logarithmic function is uniformly continuous on the support of our distributions ($[0.5, \infty)$) and $\frac{d}{du}[uK_h(u)]$ is bounded for any $h \in (0,1]$ on the interval $[0,h]$, we can conclude that the whole integrand in Eq. 28 converges in $\mathcal{L}^1$, indicating that

$$\lim_{n \to \infty} \hat{\xi}_{h,n}^{\text{pos}} = \int_{[0,h]} \ln Q(1-u) \frac{d}{du}[uK_h(u)] \, du \text{ with probability } 1. \tag{29}$$

We note that when the terms $\hat{q}_{h,n}^{(1)}$ and $\hat{q}_{h,n}^{(1)}$ are converted to a form similar to Eq. 28, the derivatives $\frac{d}{du} u^\gamma K_h(u)$ and $\frac{d^2}{du^2} u^{\gamma+1} K_h(u)$ inside the integral are not bounded if $0.5 < \gamma < 1$ due to the term $u^{\gamma-1}$ diverging at zero. However, we show that even in this case the interchange of limit and integral is allowed (see Proposition 1 at the end of this section; the proof applies to any interval $[0,h], h \in (0,1]$). Note that the proof uses the assumption that the $\mathbb{E}[X^2]$ is finite, which applies for the lognormal and the stretched exponential distributions analyzed in Section IV of the article.

As the integral of Eq. 29 is finite, it coincides with the improper Riemann integral. We obtain the value of the integral numerically. Since the PDFs in our case are step functions, the corresponding quantile function as well as the CCDF are linear piecewise functions. To obtain an analytically integrable expression, we break the integral into a sum, where the limits of the integral in each summand correspond to the breakpoints of the quantile function $Q$. Since $\ln Q(1-u)$ is differentiable between these limits, the integral can be converted to a sum of Riemann integrals:

$$\xi_h^{\text{pos}} = -\sum \int_a^b uK_h(u) \frac{d}{du}[\ln Q(1-u)] \, du \tag{30}$$

Denoting the slope of $Q$ between breakpoints $a$ and $b$ with $m$ and the intercept with $c$, we obtain:

$$\xi_h^{\text{pos}} = -\sum \int_a^b uK_h(u) \frac{m}{c - mu + m} \, du \tag{31}$$

We solve this integral inside the summation analytically, loop over the summands and find the variables $m$ and $c$ for each summand the way we would determine the equation of any linear graph; if $a$ and $b$ denote the limits of the integral, the slope is given by $(Q(1-b) - Q(1-a))/((1-b) - (1-a))$, and the intercept of the equation is solved accordingly. The values of $\hat{q}_h^{(1)}$ and $\hat{q}_h^{(2)}$ can be computed in a similar fashion.

While the above method is fairly fast and scales well, we find it to be very sensitive to the estimation of $m$, and for certain types of distributions the estimate starts to oscillate in an undesirable fashion when moving towards the right endpoint of the distribution. In these cases, we calculate $\xi_h^{\text{pos}}$ as

$$\xi_h^{\text{pos}} = \sum \int_a^b \ln Q(1-u) \frac{d}{du}[uK_h(u)] \, du \tag{32}$$

using the *quad* function of *scipy.integrate* to obtain the values of the integrals in each summand. This method, however, scales poorly as the number of summands increases. To speed up the integration, we perform the actual integration in C via the Python library *ctypes*.

We use these two methods in a complementary fashion; most of the time we use the first method due to its speed, but in case of doubt, the results are compared to those obtained with the second method.

**Lemma 1.** *Let $X_1, X_2, \ldots$ be independent $F$-distributed random variables. Assume that $\mathbb{E}|X_1|^p < \infty$ for some $1 < p < \infty$. Then for any $\gamma > \frac{1}{p} - 1$,*

$$\mathbb{E} \int_0^1 u^\gamma |Q_n(1-u) - Q(1-u)| \, du \leq c \left(\mathbb{E} W_p^p(F_n, F)\right)^{1/p},$$

*where $c$ is a finite constant and $W_p$ indicates the Wasserstein distance of order $p$ between probability measures corresponding to the empirical cumulative distribution $F_n$ and the cumulative distribution function $F$.*

*Proof.* Denote

$$\Delta_n = \int_0^1 u^\gamma |Q_n(1-u) - Q(1-u)| \, du.$$

Define $q > 1$ by the formula $\frac{1}{q} = 1 - \frac{1}{p}$. Hölder's inequality then implies that

$$\Delta_n \leq \left(\int_0^1 u^{\gamma q}\, du\right)^{1/q} \left(\int_0^1 |Q_n(1-u) - Q(1-u)|^p\, du\right)^{1/p}$$

$$= c \left(\int_0^1 |Q_n(u) - Q(u)|^p\, du\right)^{1/p},$$

where the constant $c = (\int_0^1 u^{\gamma q}\, du)^{1/q}$ is finite because the assumption $\gamma > \frac{1}{p} - 1 = -\frac{1}{q}$ implies that $\gamma q > -1$. In particular,

$$\mathbb{E}\Delta_n^p \leq c^p \, \mathbb{E} \int_0^1 |Q_n(u) - Q(u)|^p\, du.$$

Then by Jensen's inequality,

$$\mathbb{E}\Delta_n \leq (\mathbb{E}\Delta_n^p)^{1/p} \leq c \left(\mathbb{E} \int_0^1 |Q_n(u) - Q(u)|^p\, du\right)^{1/p}.$$

Next, Theorem 4.4 of Ref. [9] implies that $\mathbb{E} \int_0^1 |Q_n(u) - Q(u)|^p\, du = \mathbb{E}W_p^p(F_n, F)$, where $W_p$ indicates the Wasserstein distance of order $p$ between the empirical CDF $F_n$ and the CDF $F$. Hence the claim follows. $\square$

**Proposition 1.** *Assume that $P(X_1 \geq t_0) = 1$ for some $t_0 > 0$, and that $\mathbb{E}X_1^p < \infty$ for some $1 < p < \infty$. Then for any $\gamma > \frac{1}{p} - 1$,*

$$\mathbb{E} \int_0^1 u^\gamma \, |\log Q_n(1-u) - \log Q(1-u)|\, du \to 0,$$

*and in particular,*

$$\int_0^1 u^\gamma \, |\log Q_n(1-u) - \log Q(1-u)|\, du \xrightarrow{\mathbb{P}} 0.$$

*Proof.* We note $F(t) = 0$ for all $t < t_0$. Therefore for any $u \in (0,1)$, $Q(u) = \inf\{t : F(t) \geq u\} = \inf\{t \geq t_0 : F(t) \geq u\} \geq t_0$. When $X_1, \ldots, X_n$ are independently sampled from $F$, it follows that $X_i \geq t_0$ for all $i$ almost surely. Therefore, a.s., $F_n(t) = 0$ for all $t < t_0$, and $Q_n(u) \geq t_0$ for all $u \in (0,1)$ a.s. Because the derivative of $\log t$ is bounded by $|(\log t)'| \leq t_0^{-1}$ on $[t_0, \infty)$, we see that $|\log y - \log x| \leq t_0^{-1}|y-x|$ for all $x, y \geq t_0$. In particular, a.s.,

$$|\log Q_n(1-u) - \log Q(1-u)| \leq t_0^{-1}|Q_n(1-u) - Q(1-u)| \qquad \text{for all } u \in (0,1).$$

It follows by the help of Lemma 1 that

$$\mathbb{E} \int_0^1 u^\gamma \, |\log Q_n(1-u) - \log Q(1-u)|\, du$$

$$\leq t_0^{-1} \mathbb{E} \int_0^1 u^\gamma \, |Q_n(1-u) - Q(1-u)|\, du$$

$$\leq c t_0^{-1} \left(\mathbb{E} W_p^p(F_n, F)\right)^{1/p}.$$

Furthermore, Theorem 2.14 of Ref. [9] implies that $\mathbb{E}W_p^p(F_n, F) \to 0$. Hence the first claim follows. The second claim follows immediately from the first by Markov's inequality. $\square$

## V. NUMERICALLY CALCULATED LIMIT VALUES OF THE ESTIMATORS

Figures 8 and 9 show the numerically obtained results regarding the limit values of the estimators for other tested parametrizations of the lognormal and the stretched exponential distribution not shown in Figures 7 and 8 of the article. The results are obtained as described in Section IV of the article.
.

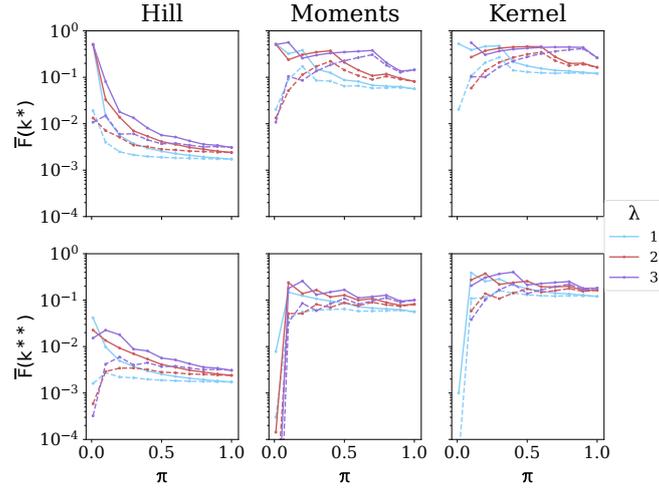

FIG. 8. Value of the CCDF at the smallest $k_{\min}$ at which the limit value of the estimators falls below $1/4$ for the first time (top row in each subfigure) and permanently (bottom row in each subfigure) for the stretched exponential distribution with $\beta = 0.5$. The value of the CCDF is expressed both relative to the subsampled distribution (solid lines) and to the original distribution (dashed lines). Note that for the stretched exponential distribution with $\lambda \in \{2, 3\}$, we do not obtain values for $\pi = 0.01$ due to unstable behaviour of the algorithm on the subsampled distributions.

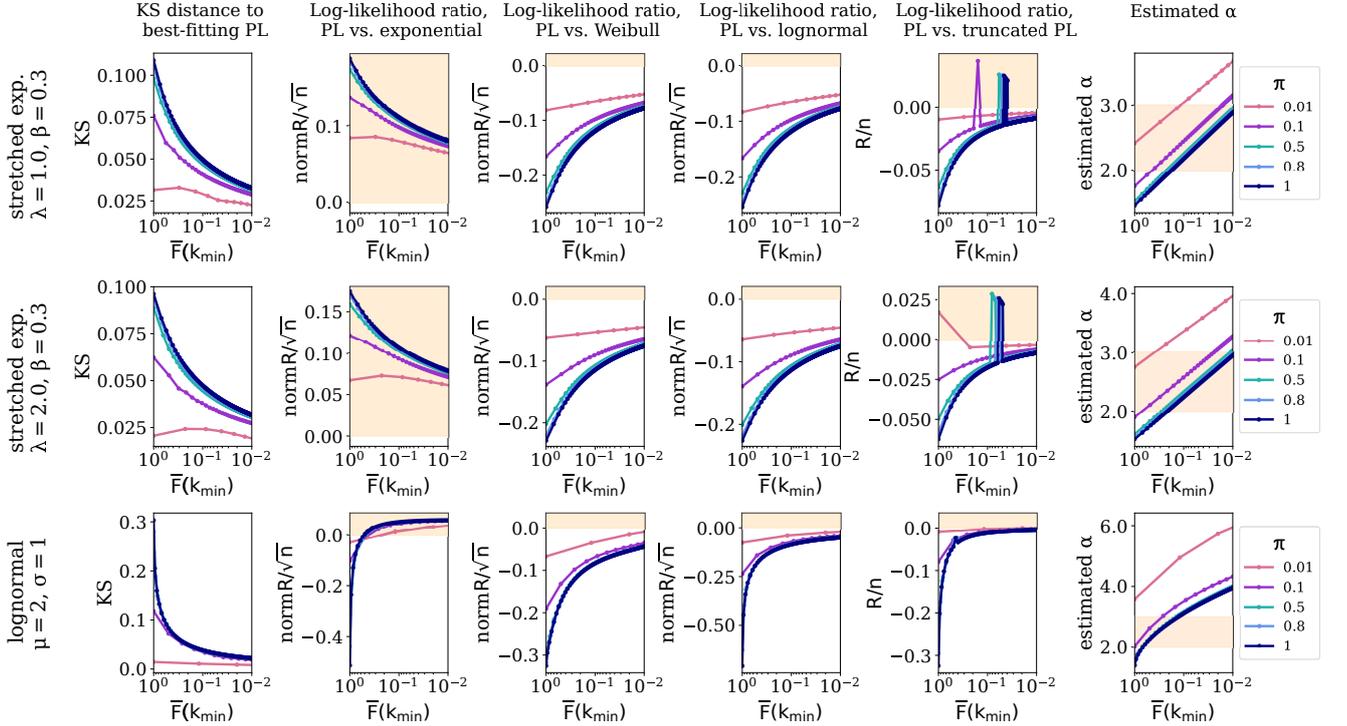

FIG. 9. Limit values of the estimators of the ML method for lognormal and stretched exponential distributions as a function of $\bar{F}(k_{\min})$ (see Fig. 8 of the article for detailed explanation).



\end{document}